\documentclass[printer]{tCPH2e}

\usepackage{graphicx}
\usepackage{rotating}

\newcommand{\sbar}[1]{\overline{#1}}
\usepackage{bar}

\begin{document}
 \doi{10.1080/0010751YYxxxxxxxx}
 \issn{1366-5812}
\issnp{0010-7514} \jvol{00} \jnum{00} \jyear{2006} 

\markboth{Roberto Trotta}{Bayes in the sky}

\title{Bayes in the sky: \\ Bayesian inference and model selection in cosmology }

\author{Roberto Trotta$^{\ast}$\thanks{$^\ast$Email:
rxt@astro.ox.ac.uk}
\\\vspace{6pt}  Oxford University, Astrophysics Department \\ Denys Wilkinson Building, Keble Rd, Oxford, OX1 3RH, UK\\
\received{\today} }

\maketitle

\begin{abstract}

The application of Bayesian methods in cosmology and astrophysics
has flourished over the past decade, spurred by data sets of
increasing size and complexity. In many respects, Bayesian methods
have proven to be vastly superior to more traditional statistical
tools, offering the advantage of higher efficiency and of a
consistent conceptual basis for dealing with the problem of
induction in the presence of uncertainty. This trend is likely to
continue in the future, when the way we collect, manipulate and
analyse observations and compare them with theoretical models will
assume an even more central role in cosmology.

This review is an introduction to Bayesian methods in cosmology
and astrophysics and recent results in the field. I first present
Bayesian probability theory and its conceptual underpinnings,
Bayes' Theorem and the role of priors. I discuss the problem of
parameter inference and its general solution, along with numerical
techniques such as Monte Carlo Markov Chain methods. I then review
the theory and application of Bayesian model comparison,
discussing the notions of Bayesian evidence and effective model
complexity, and how to compute and interpret those quantities.
Recent developments in cosmological parameter extraction and
Bayesian cosmological model building are summarized, highlighting
the challenges that lie ahead.

\medskip

\end{abstract}

\newcommand{\params}{{\theta}}
\newcommand{\data}{{d}}
\newcommand{\mdl}{{\mathcal M}}
\newcommand{\like}{{\mathcal L}}
\newcommand{\mlike}{{\mathcal E}}
\newcommand{\Ord}[1]{{\mathcal O}(#1)}
\newcommand{\dchisq}{\Delta \chi^2}
\newcommand\lsim{\mathrel{\rlap{\lower4pt\hbox{\hskip1pt$\sim$}}
    \raise1pt\hbox{$<$}}}
\newcommand\gsim{\mathrel{\rlap{\lower4pt\hbox{\hskip1pt$\sim$}}
    \raise1pt\hbox{$>$}}}
\newcommand{\dr}{{\rm d}}
\newcommand{\pzero}{{\phi}}
\newcommand{\pone}{{\psi}}
\newcommand{\LL}{{\mathcal{L}}}
\newcommand{\vc}[1]{{{#1}}}
\newcommand{\KL}{D_{{\rm KL}}}
\newcommand{\CC}{{\cal C}}
\newcommand{\Otot}{\Omega_{\rm{tot}}}
\newcommand{\p}{{\wp}}
\newcommand{\bB}{{\overline{B}_{10}}}
\newcommand{\lmax}{{\cal L}_{\rm max}}
\newcommand{\pml}{\params_{\rm max}}
\newcommand{\chisq}{\chi^2}
\newcommand{\It}{I_{10}}
\newcommand{\LCDM}{$\Lambda$CDM}
\newcommand{\weff}{w_{\rm{eff}}}
\newcommand{\seff}{\sigma_{\rm{eff}}}
\newcommand{\df}{e}
\newcommand{\pfid}{\psi^\star}
\newcommand{\pt}{\tilde{p}}

\newcommand{\be}{\begin{equation}}
\newcommand{\ee}{\end{equation}}
\newcommand{\note}[1]{{\bf #1}}
\newcommand{\dtab}[1]{\multicolumn{3}{l}{{\bf #1}} \\}
\newcommand{\range}[2]{$\left[#1, #2\right]$}
\newcommand{\define}[1]{\\\vspace{\baselineskip}``{\em #1
}''}\vspace{\baselineskip}
\newcommand{\preprint}[1]{Available as pre--print: #1}

\newcommand{\ct}[1]{#1}
\newcommand{\xcite}[2]{\cite{#1,#2}}
\newcommand{\ncite}[1]{\cite{#1}}

\newcommand{\eqref}[1]{(\ref{#1})}
\newcommand{\eg}{e.g.}
\newcommand{\ie}{i.e.}
\newcommand{\update}{{\bf update}}
\newcommand{\yr}[1]{{\rule{10pt}{0pt}\small\textbf{#1}}}
\newcommand{\mbar}[3]{\sethspace{0.0}
\bar{#2}{8}[\yr{#3}]\sethspace{0.5}\bar{#1}{4}}
\newcommand{\cosmomc}{\texttt{CosmoMC}}

\def\Ob{\Omega_b}
\def\Oc{\Omega_{\rm cdm}}
\newcommand{\Ocrit}{\Omega_{\rm crit}}
\def\Ok{\Omega_{\kappa}}
\def\Ol{\Omega_\Lambda}
\newcommand{\Og}{\Omega_\gamma}
\newcommand{\Ode}{\Omega_{\rm de}}
\def\Om{\Omega_m}
\def\Od{\Omega_d}
\def\On{\Omega_\nu}
\def\Or{\Omega_r}
\def\ob{\omega_b}
\def\oc{\omega_{\rm c}}
\def\ocdm{\omega_{\rm c}}
\def\od{\omega_d}
\def\ok{\omega_k}
\def\ol{\omega_\Lambda}
\def\om{\omega_{\rm m}}
\def\on{\omega_\nu}
\def\fb{f_b}
\def\fn{f_\nu}
\def\ns{{n_s}}
\def\nt{{n_t}}
\def\al{\alpha}
\def\Ot{\Omega_{\rm tot}}
\def\As{A_s}
\def\At{A_t}
\def\Ap{A_{\rm peak}}
\def\Apivot{A_{\rm pivot}}
\def\zeq{{z_{\rm eq}}}
\def\zrec{{z_{\rm rec}}}
\def\zion{{z_{\rm ion}}}
\def\zacc{{z_{\rm acc}}}
\def\teq{t_{\rm eq}}
\def\trec{t_{\rm req}}
\def\tion{t_{\rm ion}}
\def\tacc{t_{\rm acc}}
\def\tnow{t_{\rm now}}
\def\age{\tnow}
\def\Th{\Theta_s}
\def\lA{\l_A}
\def\dA{d_A}
\def\dV{d_V}
\def\Mnu{M_\nu}
\def\Nnu{N_\nu}
\def\Qnl{Q_{\rm nl}}
\def\Op{A_\tau}
\def\kstar{k_*}
\def\Tcmb{T_{\rm cmb}}

\section{Introduction}

At first glance, it might appear surprising that a trivial
mathematical result obtained by an obscure minister over 200
hundred years ago ought still to excite so much interest across so
many disciplines, from econometrics to biostatistics, from
financial risk analysis to cosmology. Published posthumously
thanks to Richard Price in 1763, ``An essay towards solving a
problem in the doctrine of chances'' by the rev.~Thomas Bayes
(1701(?)--1761)~\ct{\cite{Bayes:1763}} had nothing in it that
could herald the growing importance and enormous domain of
application that the subject of {\em Bayesian probability theory}
would acquire more than two centuries afterwards. However, upon
reflection there is a very good reason why Bayesian methods are
undoubtedly on the rise in this particular historical epoch: the
exponential increase in computational power of the last few
decades made massive numerical inference feasible for the first
time, thus opening the door to the exploitation of the power and
flexibility of a rich set of Bayesian tools. Thanks to fast and
cheap computing machines, previously unsolvable inference problems
became tractable, and algorithms for numerical simulation
flourished almost overnight.

\begin{figure}
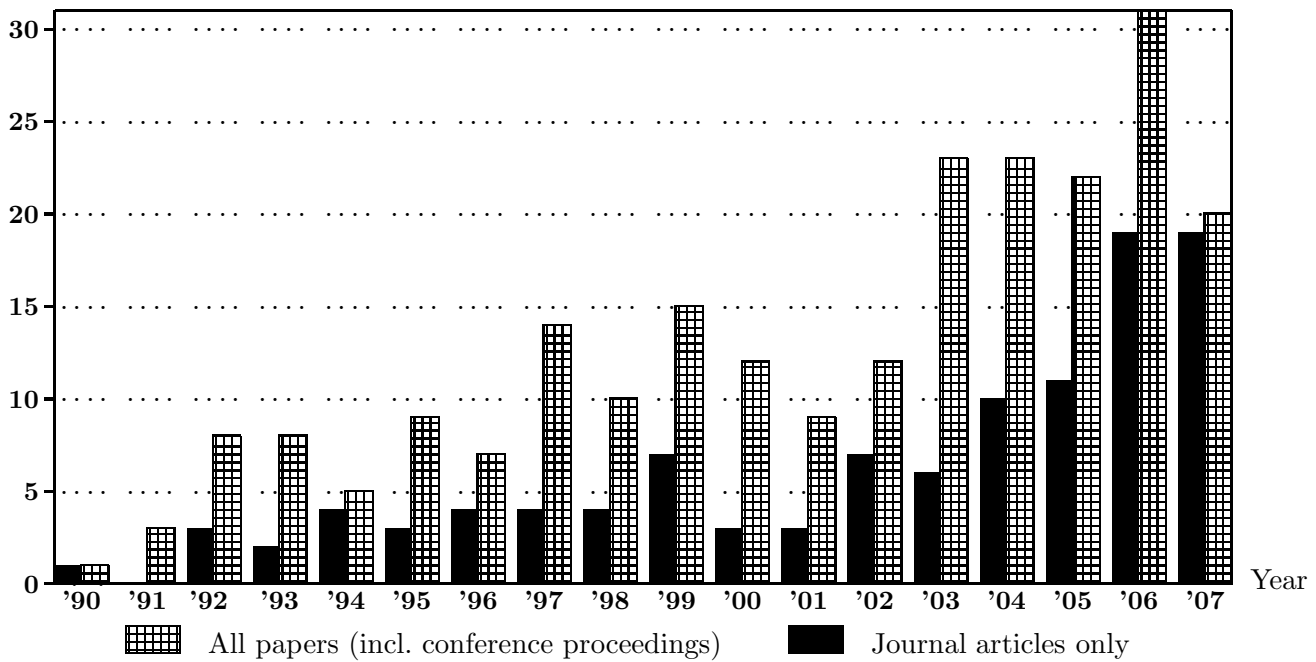

\begin{center}
\textbf{Number of Bayesian papers in cosmology and astrophysics}\\[5mm]
\begin{barenv}
\setwidth{10} \setdepth{0} \setstretch{7} \setnumberpos{side}
 \setxname{\raisebox{+5ex}{~~~~~~~~~~Year}}
  \setstyle{\small\bfseries}
  \setyaxis{0}{31}{5}
   \setstyle{\small}
   \hlineon
   \mbar{1}{1}{'90}
   \mbar{3}{0}{'91}
   \mbar{8}{3}{'92 }
   \mbar{8}{2}{'93}
   \mbar{5}{4}{'94}
   \mbar{9}{3}{'95}
   \mbar{7}{4}{'96}
   \mbar{14}{4}{'97}
   \mbar{10}{4}{'98}
   \mbar{15}{7}{'99}
   \mbar{12}{3}{'00}
   \mbar{9}{3}{'01}
   \mbar{12}{7}{'02}
   \mbar{23}{6}{'03}
   \mbar{23}{10}{'04}
   \mbar{22}{11}{'05}
   \mbar{31}{19}{'06}
   \mbar{20}{19}{'07}
\end{barenv}
\par\vspace{1\baselineskip}
\legend{4}{All papers (incl. conference proceedings)} \qquad
\legend{8}{Journal articles only}
\end{center}
\caption{The evolution of the B--word: number of articles in
astronomy and cosmology with ``Bayesian'' in the title, as a
function of publication year. The number of papers employing one
form or another of Bayesian methods is of course much larger than
that. Up until about 1995, Bayesian papers were concerned mostly
with image reconstruction techniques, while in subsequent years
the domain of application grew to include signal processing,
parameter extraction, object detection, cosmological model
building, decision theory and experiment optimization, and much
more. It appears that interest in Bayesian statistics began
growing around 2002~(source: NASA/ADS).
\label{fig:Bayesian_papers}}
\end{figure}

Historically, the connections between physics and Bayesian
statistics have always been very strong. Many ideas were developed
because of related physical problems, and physicists made several
distinguished contributions. One has only to think of people like
Laplace, Bernouilli, Gauss, Metropolis, Jeffreys, etc. Cosmology
is perhaps among the latest disciplines to have embraced Bayesian
methods, a development mainly driven by the data explosion of the
last decade, as Figure~\ref{fig:Bayesian_papers} indicates.
However, motivated by difficult and computationally intensive
inference problems, cosmologists are increasingly coming up with
new solutions that add to the richness of a growing Bayesian
literature.

Some cosmologists are sceptic regarding the usefulness of
employing more advanced statistical methods, perhaps because they
think with Mark Twain that there are ``lies, damned lies and
statistics''. One argument that is often heard is that there is no
point in bothering too much about refined statistical analyses, as
better data will in the future resolve the question one way or
another, be it the nature of dark energy or the initial conditions
of the Universe. I strongly disagree with this view, and would
instead argue that sophisticated statistical tools will be
increasingly central for
 modern cosmology. This opinion is motivated by the following reasons: \\
 \begin{enumerate}
 \item  The complexity of the modelling of both our theories and
 observations will always increase, thus requiring correspondingly more refined statistical
 and data analysis skills. In fact, the scientific return of the next generation of
 surveys will be limited by the level of sophistication and
 efficiency of our inference tools.
 \item The discovery zone for new physics is when a potentially new effect
 is seen at the 3--4 $\sigma$ level. This is when tantalizing suggestion for an effect
 starts to accumulate but there is no firm evidence yet. In this potential discovery region a careful application
 of statistics can make the difference between claiming or missing a
 new discovery.
 \item If you are a theoretician, you do not want to
 waste your time trying to explain an effect that is not there in
 the first place. A better appreciation of the interpretation of
 statistical statements might help in identifying robust claims
 from spurious ones.
 \item Limited resources mean that we need to focus our efforts on
 the most promising avenues. Experiment forecast and optimization
 will increasingly become prominent as we need to use all of our
 current knowledge ({\em and} the associated uncertainty) to
 identify the observations and strategies that are likely to give the highest
 scientific return in a given field.
 \item Sometimes there will be no better data! This is the case
 for the many problems associated with cosmic variance limited
 measurements on large scales, for example in the cosmic
 background radiation, where the small number of independent
 directions on the sky makes it impossible to reduce the error
 below a certain level.
 \end{enumerate}

This review focuses on Bayesian methodologies and related issues,
presenting some illustrative results where appropriate and
reviewing the current state--of--the art of Bayesian methods in
cosmology. The emphasis is on the innovative character of Bayesian
tools. The level is introductory, pitched for graduate students
who are approaching the field for the first time, aiming at
bridging the gap between basic textbook examples and application
to current research. In the last sections we present some more
advanced material that we hope might be useful for the seasoned
practitioner, too. A basic understanding of cosmology and of the
interplay between theory and cosmological observations (at the
level of the introductory chapters in~\cite{Dodelson:2003}) is
assumed. 
A full list of references is provided as a comprehensive guidance
to relevant literature across disciplines.

This paper is organized in two main parts. The first part,
sections~\ref{sec:prob_theory}--\ref{sec:modcomp}, focuses on
probability theory, methodological issues and Bayesian methods
generally. In section~\ref{sec:prob_theory} we present the
fundamental distinction between probability as frequency or as
degree of belief, we introduce Bayes' Theorem and discuss the
meaning and role of priors in Bayesian theory.
Section~\ref{sec:Bayesian_parameter_inference} is devoted to
Bayesian parameter inference and related issues in parameter
extraction. Section~\ref{sec:modcomp} deals with the topic of
Bayesian model comparison from a conceptual and technical point of
view, covering Occam's razor principle, its practical
implementation in the form of the Bayesian evidence, the effective
number of model parameters and information criteria for
approximate model comparison. The second part presents
applications to cosmological parameter inference and related
topics (section~\ref{sec:cosmo_par_inferece}) and to Bayesian
cosmological model building (section
\ref{sec:cosmo_model_building}), including multi--model inference
and model comparison forecasting. Section~\ref{sec:conclusion}
gives our conclusions.

\section{Bayesian probability theory}
\label{sec:prob_theory}

In this section we introduce the basic concepts and the notation
we employ. After a discussion of what probability is, we turn to
the central formula for Bayesian inference, namely Bayes theorem.
The whole of Bayesian inference follows from this extremely simple
cornerstone. We then present some views about the meaning of
priors and their role in Bayesian theory, an issue which has
always been (wrongly) considered a weak point of Bayesian
statistics.

There are many excellent textbooks on Bayesian statistics: the
works by Sir Harold Jeffreys~\cite{Jeffreys:1961} and Bruno de
Finetti~\cite{deFinetti:1974} are classics, while an excellent
modern introduction with an extensive reading list is given
by~\cite{Bernardo:1994}. A good textbook is \cite{Box:1992}. Worth
reading as a source of inspiration is the though--provoking
monograph by E.T.~Jaynes~\cite{JaynesBook}. Computational aspects
are treated in~\cite{Marin:2007}, while MacKay~\cite{MacKay:2003}
has a quite unconventional but inspiring choice of topics with
many useful exercices. Two very good textbooks on the subject
written by physicists are~\cite{Sivia:1996,Gregory:2005}. A nice
introductory review aimed at physicists
is~\cite{D'Agostini:1995fv} \ct{(see
also~\cite{D'Agostini:2003qr})}. Tom Loredo has some masterfully
written introductory material,
too~\xcite{Loredo:1990}{Loredo:1992}. A good source expanding on
many of the topics covered here is Ref.~\cite{Liddle:BMIC2008}.

\subsection{What is probability?}

\subsubsection{Probability as frequency}

The classical approach to statistics defines the probability of an
event as
 \begin{quote}
 {\em ``the number of times the event occurs over the total number of
trials, in the limit of an infinite series of equiprobable
repetitions.''}
\end{quote}
This is the so--called {\em frequentist} school of thought. This
definition of probability is however unsatisfactory in many
respects.
\begin{enumerate}
 \item Strikingly, this definition of probability in terms of
 relative frequency of outcomes is {\em circular}, i.e.~it assumes that
 repeated trials have the same probability
 of outcomes -- but it was the the very notion of {\em probability} that we were
 trying to define in the first place!
 \item It cannot handle with unrepeatable
situations, such as the probability that I will be overrun by a
car when crossing the street, or, in the cosmological context,
questions concerning the properties of the observable Universe as
a whole, of which we have exactly one sample. Indeed, perfectly
legitimate questions such as ``what is the probability that it was
raining in Oxford when William I was crowned?'' cannot even be
formulated in classical statistics.
 \item The definition only holds
 exactly for an infinite sequence of repetitions. In practice we
 always handle with a finite number of measurements, sometimes
 with actually only a very small number of them. How can we assess
 when ``how many repetitions'' are sufficient? And what shall we do when
 we have only a handful of repetitions? Frequentist statistics does
 not say, except sometimes devising complicated {\em ad-hockeries} to correct for ``small
 sample size'' effects. In practice, physicists tend to forget about the
 ``infinite series'' requirement and use this definitions and the results that go with it
 (for example, about asymptotic distributions of test statistics)
 for whatever number of samples they happen to be working with.
\end{enumerate}

Another, more subtle aspects has to do with the notion of
``randomness''. Restricting ourselves to classical (non--chaotic)
physical systems for now, let us consider the paradigmatic example
of a series of coin tosses. From an observed sequence of heads and
tails we would like to come up with a statistical statement about
the fairness of the coin, which is deemed to be ``fair'' if the
probability of getting heads is $p_H=0.5$. At first sight, it
might appear plausible that the task is to determine whether the
coin possesses some physical property (for example, a tensor of
intertia symmetric about the plane of the coin) that will ensure
that the outcome is indifferent with respect to the interchange of
heads and tails. As forcefully argued by Jaynes~\cite{JaynesBook},
however, the probability of the outcome of a sequence of tosses
has nothing to do with the physical properties of the coin being
tested! In fact, a skilled coin--tosser \ct{(or a purpose--built
machine, see~\cite{Diaconis:2007})} can influence the outcome
quite independently of whether the coin is well--balanced (i.e.,
symmetric) or heavily loaded. The key to the outcome is in fact
the definition of {\em random} toss. In a loose, intuitive
fashion, we sense that a carefully controlled toss, say in which
we are able to set quite precisely the spin and speed of the coin,
will spoil the ``randomness'' of the experiment --- in fact, we
might well call it ``cheating''. However, lacking a precise
operational definition of what a ``random toss'' means, we cannot
meaningfully talk of the probability of getting heads as of a
physical property of the coin itself. It appears that the outcome
depends on our {\em state of knowledge} about the initial
conditions of the system (angular momentum and velocity of the
toss): an lack of precise information about the initial conditions
results in a state of knowledge of indifference about the possible
outcome with respect to the specification of heads or tails. If
however we insist on defining probability in terms of the outcome
of random experiments, we immediately get locked up in a
circularity when we try to specify what ``random'' means. For
example, one could say that
 \begin{quote}
 {\em ``a random toss is one for which the sequence of heads and tails is
compatible with assuming the hypothesis $p_H = 0.5$''}.
 \end{quote}
But the latter statement is exactly what we were trying to test in
the first place by using a sequence of random tosses! We are back
to the problem of circular definition we highlighted above.

\subsubsection{Probability as degree of belief}

Many of the limitations above can be avoided and paradoxes
resolved by taking a Bayesian stance about probabilities. The
Bayesian viewpoint is based on the simple and intuitive {\em
tenet} that
 \begin{quote}
 {\em ``probability is a measure of the degree of belief about a
 proposition''.}
 \end{quote}
It is immediately clear that this definition of probability
applies to any event, regardless whether we are considering
repeated experiments (e.g., what is the probability of obtaining
10 heads in as many tosses of a coin?) or one--off situations
(e.g., what is the probability that it will rain tomorrow?).
Another advantage is that it deals with uncertainty independently
of its origin, i.e. there is no distinction between ``statistical
uncertainty'' coming from the finite precision of the measurement
apparatus and the associated random noise and ``systematic
uncertainty'', deriving from deterministic effects that are only
partially known (e.g., calibration uncertainty of a detector).
From the  coin tossing example above we learn that it makes good
sense to think of probability as a state of knowledge in presence
of partial information and that ``randomness'' is really a
consequence of our lack of information about the exact conditions
of the system (if we knew the precise way the coin is flipped we
could predict the outcome of any toss with certainty. The case of
quantum probabilities is discussed below). The rule for
manipulating states of belief is given by Bayes' Theorem, which is
introduced in Eq.~\eqref{eq:Bayes_Theorem_hypothesis} below.

It seems to us that the above arguments strongly favour the
Bayesian view of probability (a more detailed discussion can be
found in~\xcite{JaynesBook}{Loredo:1990}). Ultimately, as
physicists we might as well take the pragmatic view that the
approach that yields demonstrably superior results ought to be
preferred. In many real--life cases, there are several good
reasons to prefer a
Bayesian viewpoint:\\
\begin{enumerate}
 \item Classic frequentist methods are often based on asymptotic properties of
estimators. Only a handful of cases exist that are simple enough
to be amenable to analytic treatment (in physical problems one
most often encounters the Normal and the Poisson distribution).
Often, methods based on such distributions are employed not
because they accurately describe the problem at hand, but because
of the lack of better tools. This can lead to serious mistakes.
Bayesian inference is not concerned by such problems: it can be
shown that {\em application of Bayes' Theorem recovers frequentist
results (in the long run) for cases simple enough where such
results exist}, while remaining applicable to questions that
cannot even be asked in a frequentist context.
 \item Bayesian inference deals effortlessly with {\em nuisance parameters}.
Those are parameters that have an influence on the data but are of
no interest for us. For example, a problem commonly encountered in
astrophysics is the estimation of a signal in the presence of a
background rate
(see~\xcite{Loredo:1990,Protassov:2002sz}{D'Agostini:2004dp}). The
particles of interest might be photons, neutrinos or cosmic rays.
Measurements of the source $s$ must account for uncertainty in the
background, described by a nuisance parameter $b$. The Bayesian
procedure is straightforward: infer the joint probability of $s$
and $b$ and then integrate over the uninteresting nuisance
parameter $b$ (``marginalization'',
see~Eq.~\eqref{eq:marginalisation_continuous}). Frequentist
methods offer no simple way of dealing with nuisance parameters
(the very name derives from the difficulty of accounting for them
in classical statistics). However neglecting nuisance parameters
or fixing them to their best--fit value can result in a very
serious underestimation of the uncertainty on the parameters of
interest \ct{(see~\cite{Andreon:2003uh} for an example involving
galaxy evolution models)}.
 \item In many situations {\em prior information} is highly relevant and
omitting it would result in seriously wrong inferences. The
simplest case is when the parameters of interest have a physical
meaning that restricts their possible values: masses, count rates,
power and light intensity are examples of quantities that must be
positive. Frequentist procedures based only on the likelihood can
give best--fit estimates that are negative, and hence meaningless,
unless special care is taken (for example, constrained likelihood
methods). This often happens in the regime of small counts or low
signal to noise. The use of Bayes' Theorem ensures that relevant
prior information is accounted for in the final inference and that
physically meaningless results are weeded out from the beginning.
 \item Bayesian statistics only deals with the {\em data that were
actually observed}, while frequentist methods focus on the
distribution of possible data that have not been obtained. As a
consequence, {\em frequentist results can depend on what the
experimenter thinks about the probability of data that have not
been observed.} (this is called the ``stopping rule'' problem).
This state of affairs is obviously absurd. Our inferences should
not depend on the probability of what could have happened but
should be conditional on whatever has actually occurred. This is
built into Bayesian methods from the beginning since inferences
are by construction conditional on the observed data.

\end{enumerate}

However one looks at the question, it is fair to say that the
debate among statisticians is far from settled \ct{(for a
discussion geared for physicists, see~\cite{Cousins:1994yw})}.
Louis Lyons neatly summarized the state of affairs by saying
that~\cite{Lyons:2006}
 \begin{quote}
{\em ``Bayesians address the question everyone is interested in by
using assumptions no--one believes, while frequentists use
impeccable logic to deal with an issue of no interest to
anyone''.}
 \end{quote}

\subsubsection{Quantum probability}

Ultimately the quantum nature of the microscopic world ensures
that the fundamental meaning of probability is to be found in a
theory of quantum measurement. The fundamental problem is how to
make sense of the indeterminism brought about when the
wavefunction collapses into one or other of the eigenstates being
measured. The classic textbook view is that the probability of
each outcome is given by the square of the amplitude, the
so--called  ``Born rule''. However the question remains
--- where do quantum probabilities come from?

Recent developments have scrutinized the scenario of the
Everettian many--worlds interpretation of quantum mechanics, in
which the collapse never happens but rather the world ``splits''
in disconnected ``branches'' every time a quantum measurement is
performed. Although all outcomes actually occur, uncertainty comes
into the picture because an observer is unsure about which outcome
will occur in her branch. David Deutsch~\ct{\cite{Deutsch:1999}}
suggested to consider the problem from a decision--theoretic point
of view. He proposed to consider a formal system in which the
probabilities (``weights'') to be assigned to each quantum branch
are expressed in terms of the preferences of a rational agent
playing a quantum game. Each outcome of the game (i.e., weights
assignment) has an associated utility function that determines the
payoff of the rational agent. Being rational, the agent will act
to maximise the expectation value of her utility. The claim is
that it is possible to find a simple and plausible set or rules
for the quantum game that allow to derive unique outcomes (i.e.,
probability assignments) in agreement with the Born rule, {\em
independently} of the utility function chosen, i.e. the payoffs.
In such a way, one obtains a bridge between subjective
probabilities (the weight assignment by the rational agent) and
quantum chance (the weights attached to the branches according to
the Born rule).

This approach to the problem of quantum measurement remains highly
controversial. For an introduction and further reading, see e.g.
\xcite{Saunders:2004}{Greaves:2007} and references therein. \ct{An
interesting comparison between classical and quantum probabilities
can be found in~\cite{Hardy:2004}.}

\subsection{Bayes' Theorem}

Bayes' Theorem is a simple consequence of the axioms of
probability theory, giving the rules by which probabilities
(understood as degree of belief in propositions) should be
manipulated. As a mathematical statement it is not controversial
--- what is a matter of debate is whether it should be used as a
basis for inference and in general for dealing with uncertainty.
In the previous section we have given some arguments why we
strongly believe this to be the case. An important result is that
Bayes' Theorem can be derived from a set of basic consistency
requirements for plausible reasoning, known as Cox
axioms\ct{~\cite{Cox:1946}}. Therefore, Bayesian probability
theory can be shown to be the unique generalization of boolean
logic into a formal system to manipulate propositions in the
presence of uncertainty~\cite{JaynesBook}. In other words,
Bayesian inference is the unique generalization of logical
deduction when the available information is incomplete.

We now turn to the presentation of the actual mathematical
framework. We adopt a fairly relaxed notation, as mathematical
rigour is not the aim of this review. For a more formal
introduction, see e.g.~\cite{Jeffreys:1961,JaynesBook}.

Let us consider a proposition $A$, which could be a random
variable (e.g., the probability of obtaining 12 when rolling two
dices) or a one--off proposition (the probability that Prince
Charles will become king in 2015), and its negation,
$\overline{A}$. The {\em sum rule} reads
 \be \label{eq:sumrule}
 p(A|I) + p(\overline{A} | I) = 1,
 \ee
where the vertical bar means that the probability assignment is
conditional on assuming whatever information is given on its
right. Above, $I$ represents any relevant information that is
assumed to be true. The {\em product rule} is written as \be
\label{eq:prodrule}
 p(A,B|I) = p(A|B,I)p(B|I),
 \ee
which says that the {\em joint probability} of $A$ and $B$ equals
to the probability of $A$ given that $B$ occurs times the
probability of $B$ occurring on its own (both conditional on
information $I$). If we are interested in the probability of $B$
alone, irrespective of $A$, the sum and product rules together
imply that
  \be \label{eq:def_marginalisation}
  p(B|I) = \sum_{A} p(A,B|I),
 \ee
where the sum runs over the possible outcomes for proposition $A$.
The quantity on the left--hand--side is called {\em marginal
probability} of $B$. Since obviously $p(A,B|I) = p(B,A|I)$, the
product rule can be rewritten to give Bayes' Theorem:
 \be
 p(B|A,I) =  \frac{p(A|B,I)p(B|I)}{p(A|I)} \quad {\rm (Bayes~
 theorem)}.
 \ee
The interpretation of this simple result is more illuminating if
one replaces for $A$ the observed data $\data$ and for $B$ the
hypothesis $H$ we want to assess, obtaining
 \be \label{eq:Bayes_Theorem_hypothesis}
 p(H | \data, I) = \frac{p(\data|H, I)
 p(H|I)}{p(\data|I)}.
 \ee
On the left--hand side, $p(H | \data, I)$ is the {\em posterior
probability} of the hypothesis taking the data into account. This
is proportional to the {\em sampling distribution} of the data
$p(\data | H,I)$ assuming the hypothesis is true, times the {\em
prior probability} for the hypothesis, $p(H | I)$ (``the prior'',
conditional on whatever external information we have, $I$), which
represents our state of knowledge before seeing the data. The
sampling distribution encodes how the degree of plausibility of
the hypothesis changes when we acquire new data. Considered as a
function of the hypothesis, for fixed data (the ones that have
been observed), it is called {\em the likelihood function} and we
will often employ the shortcut notation $\like(H) \equiv p(\data |
H,I)$. Notice that as a function of the hypothesis the likelihood
is {\em not} a probability distribution. The normalization
constant on the right--hand--side in the denominator is the {\em
marginal likelihood} (in cosmology often called the ``Bayesian
evidence'') given by
 \be \label{eq:evidence_def}
 p(\data | I) \equiv \sum_{H}  p(\data | H, I)
p( H| I) \quad {\rm (Bayesian~
 evidence)}.
 \ee
where the sum runs over all the possible outcomes for the
hypothesis $H$. This is the central quantity for model comparison
purposes, and it is further discussed in
section~\ref{sec:modcomp}. The posterior is the relevant quantity
for Bayesian inference as it represents our state of belief about
the hypothesis after we have considered the information in the
data (hence the name). Notice that there is a {\em logical}
sequence in going from the prior to the posterior, not necessarily
a {\em temporal} one, i.e. a scientist might well specify the
prior after the data have been gathered provided that the prior
reflects her state of knowledge irrespective of the data. Bayes'
Theorem is therefore a prescription as to how one learns from
experience. It gives a unique rule to update one's beliefs in the
light of the observed data.

The need to specify a prior describing a ``subjective'' state of
knowledge has exposed Bayesian inference to the criticism that it
is not objective, and hence unfit for scientific reasoning.
Exactly the contrary is true --- a thorny issue to which we now
briefly turn out attention.

\subsection{Subjectivity, priors and all that}

The prior choice is a fundamental ingredient of Bayesian
statistics. Historically, it has been regarded as problematic,
since the theory does not give guidance about how the prior should
be selected. Here we argue that this issue has been given undue
emphasis and that prior specification should be regarded as a {\em
feature} of Bayesian statistics, rather than a limitation.

The guiding principle of Bayesian probability theory is that there
can be no inference without assumptions, and thus the prior choice
ought to reflect as accurately as possible one's assumptions and
state of knowledge about the problem in question before the data
come along. Far from undermining objectivity, this is obviously a
positive feature, because Bayes' Theorem gives a {\em univoque}
procedure to update different degrees of beliefs that different
scientist might have held before seeing the data. Furthermore,
there are many cases where prior (i.e., external) information is
relevant and it is sensible to include it in the inference
procedure\footnote{As argued above, often it would be a mistake
not to do so, for example when trying to estimate a mass $m$ from
some data one should enforce it to be a positive quantity by
requiring that $p(m) = 0$ for $m<0$.}.

It is only natural that two scientists might have different priors
as a consequence of their past scientific experiences, theoretical
outlook and based on the outcome of previous observations they
might have performed. As long as the prior $p(H|I)$ (where the
extra conditioning on $I$ denotes the external information of the
kind listed above) has a support that is non--zero in regions
where the likelihood is large, repeated application of Bayes
theorem, Eq.~\eqref{eq:Bayes_Theorem_hypothesis}, will lead to a
posterior pdf that converges to a common, i.e. objective inference
on the hypothesis. As an example, consider the case where the
inference concerns the value of some physical quantity $\params$,
in which case $p(\params | \data, I)$ has to be interpreted as the
posterior probability {\em density} and $p(\params | \data, I) \dr
\params$ is the probability of $\params$ to take on a value between
$\params$ and $\params + \dr \params$. Alice and Bob have
different prior beliefs regarding the possible value of $\params$,
perhaps based on previous, independent measurements of that
quantity that they have performed. Let us denote their priors by
$p(\params | I_i)$ ($i=A,B$) and let us assume that they are
described by two Gaussian distributions of mean $\mu_{i}$ and
variance $\Sigma_i^2$, $i=A,B$ representing the state of knowledge
of Alice and Bob, respectively. Alice and Bob go together in the
lab and perform a measurement of $\theta$ with an apparatus
subject to Gaussian noise of known variance $\sigma^2$. They
obtain a value $m_1$, hence their likelihood is
 \be
 \like(\theta) \equiv p(m_1|\params) = \like_0
 \exp\left(-\frac{1}{2}\frac{(\theta-m_1)^2}{\sigma^2}\right).
 \ee
Replacing the hypothesis $H$ by the continuous variable $\params$
in Bayes' Theorem\footnote{Strictly speaking, Bayes' Theorem holds
for discrete probabilities and the passage to hypotheses
represented by continuous variables ought to be performed with
some mathematical care. Here we simply appeal to the intuition of
physicists without being too much concerned by mathematical
rigour.}, we obtain for their respective posterior pdf's after the
new datum
 \be \label{eq:Bayes_Theorem_simple}
 p(\params | m_1, I_i) = \frac{\like(\params)
 p(\params | I_i)}{p(m_1|I_i)} \quad (i=A,B).
 \ee
It is easy to see that the posterior pdf's of Alice and Bob are
again Gaussians with means
 \begin{equation}
 \overline{\mu}_i = \frac{m_1 + (\sigma/\Sigma_i)^2 \mu_i}{1 +
 (\sigma/\Sigma_i)^2}
 \end{equation}
 and variance
 \begin{equation}
 \tau_i^2 = \frac{\sigma^2}{1 + (\sigma/\Sigma_i)^2} \quad (i=A,B).
 \end{equation}
Thus if the likelihood is more informative than the prior, i.e.
for $(\sigma/\Sigma_i) \ll 1$ the posterior means of Bob and Alice
will converge towards the measured value, $m_1$. As more and more
data points are gathered, one can simply replace $m_1$ in the
above equations by the mean $\sbar{m}$ of the observations and
$\sigma^2$ by $\sigma^2/N$, with $N$ the number of data points.
Thus we can see that the initial prior means $\mu_i$ of Alice and
Bob will progressively be overridden by the data. This process is
illustrated in Figure~\ref{fig:prior_to_posterior}.

\newcommand{\spazio}{\hspace{0.21\linewidth}}
\begin{figure}[tb]
(a) \spazio (b) \spazio (c) \spazio (d) \\
\centering
\includegraphics[width=0.24\linewidth]{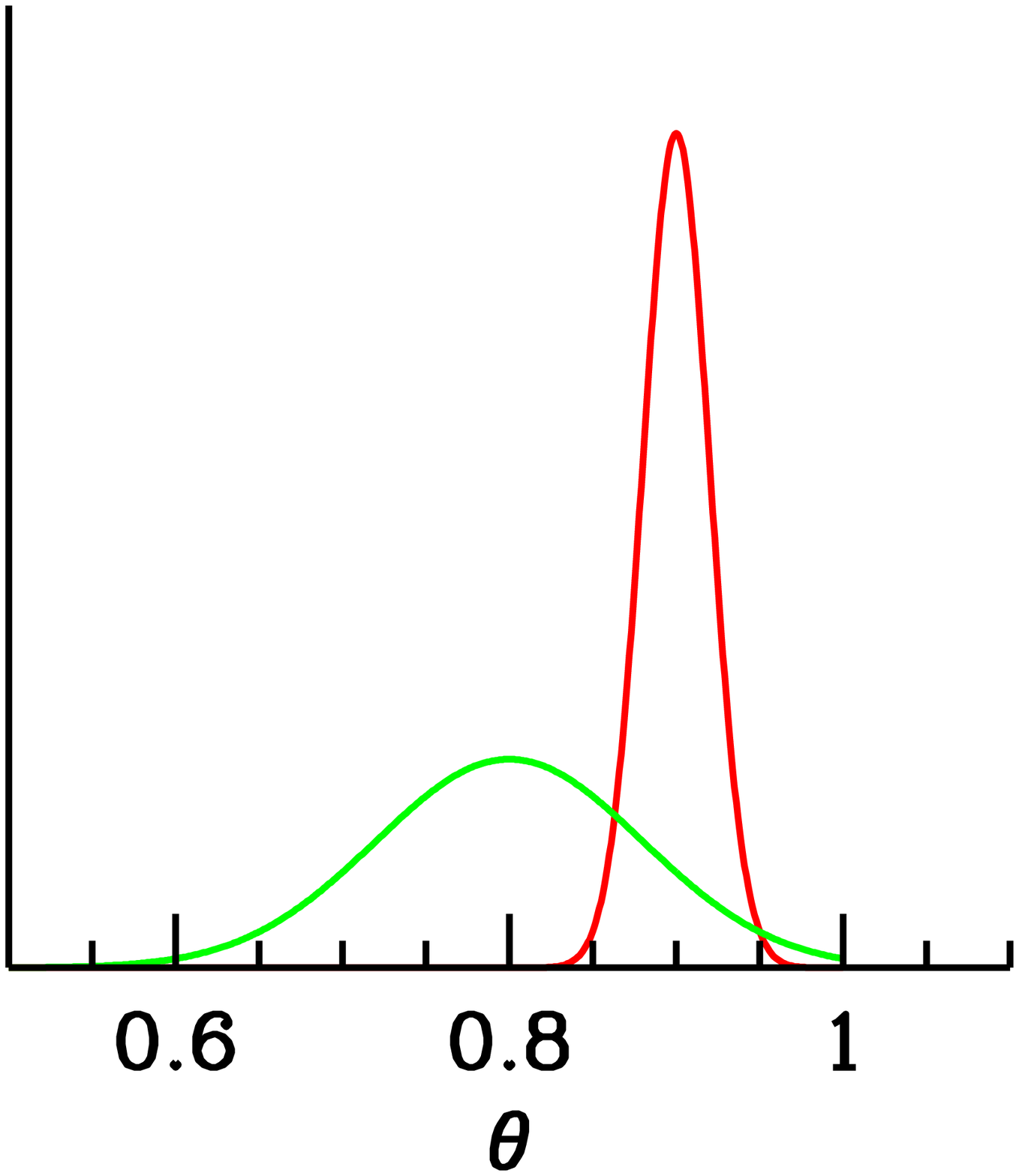}\hfill
\includegraphics[width=0.24\linewidth]{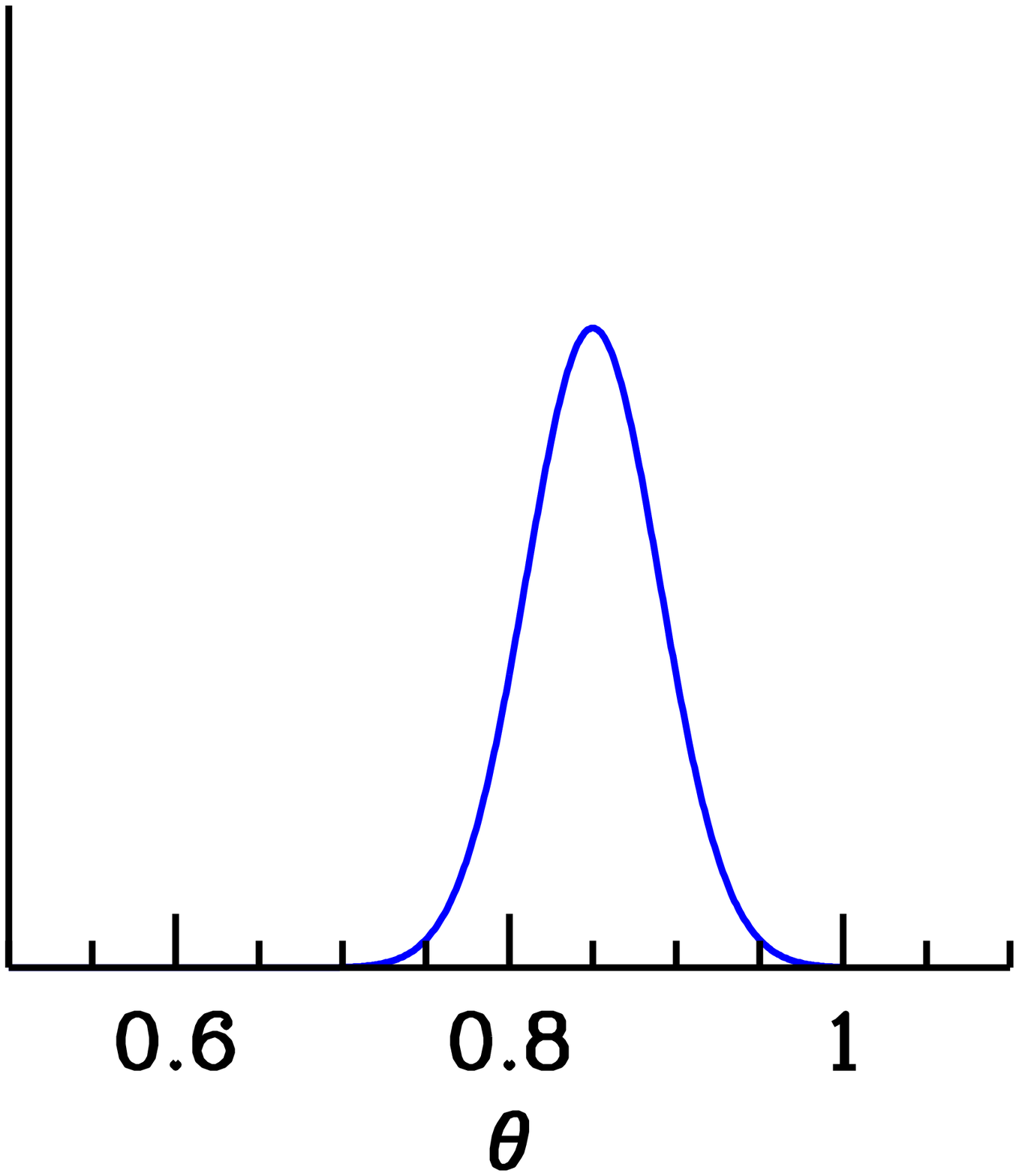}\hfill
\includegraphics[width=0.24\linewidth]{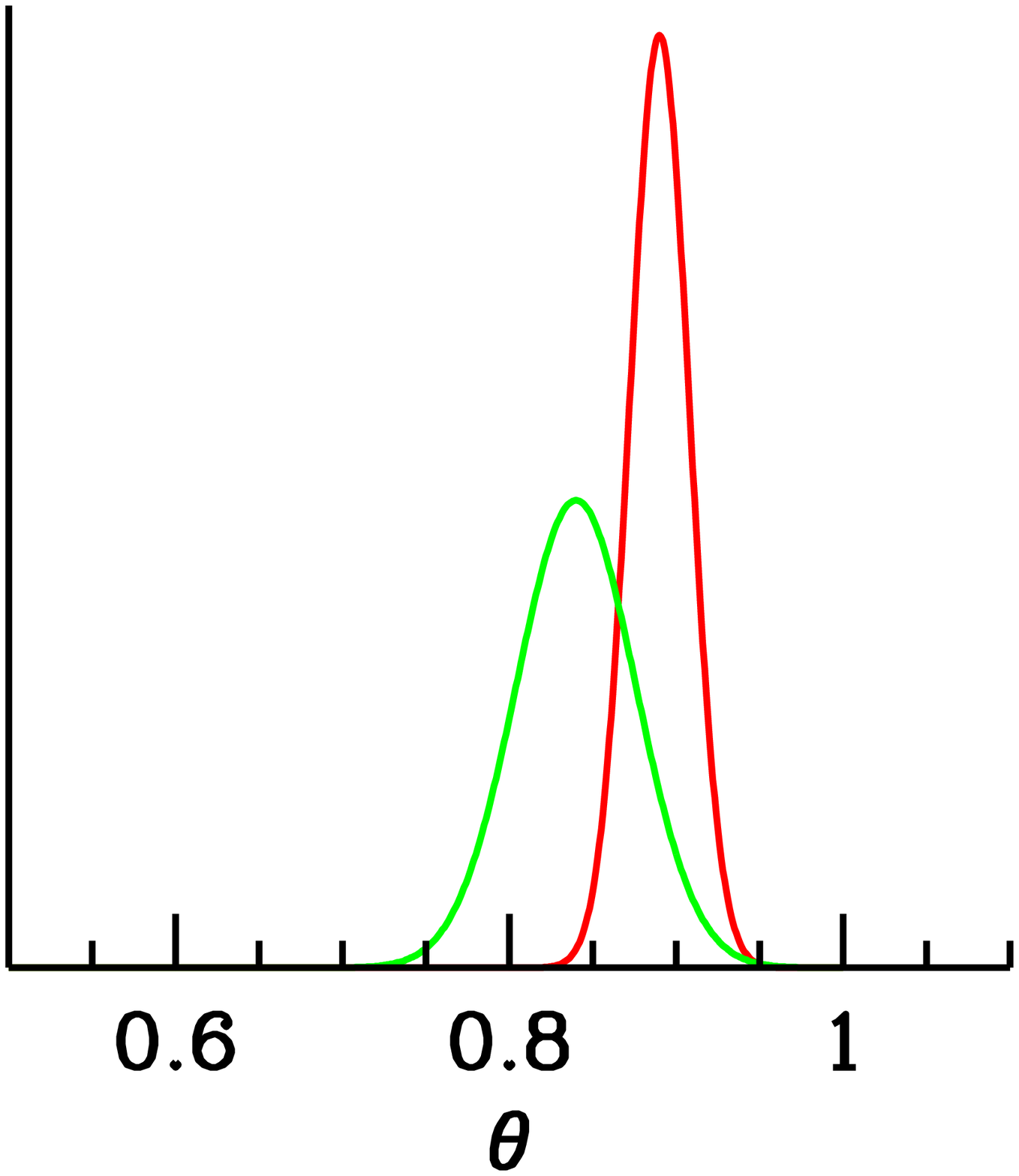}\hfill
\includegraphics[width=0.24\linewidth]{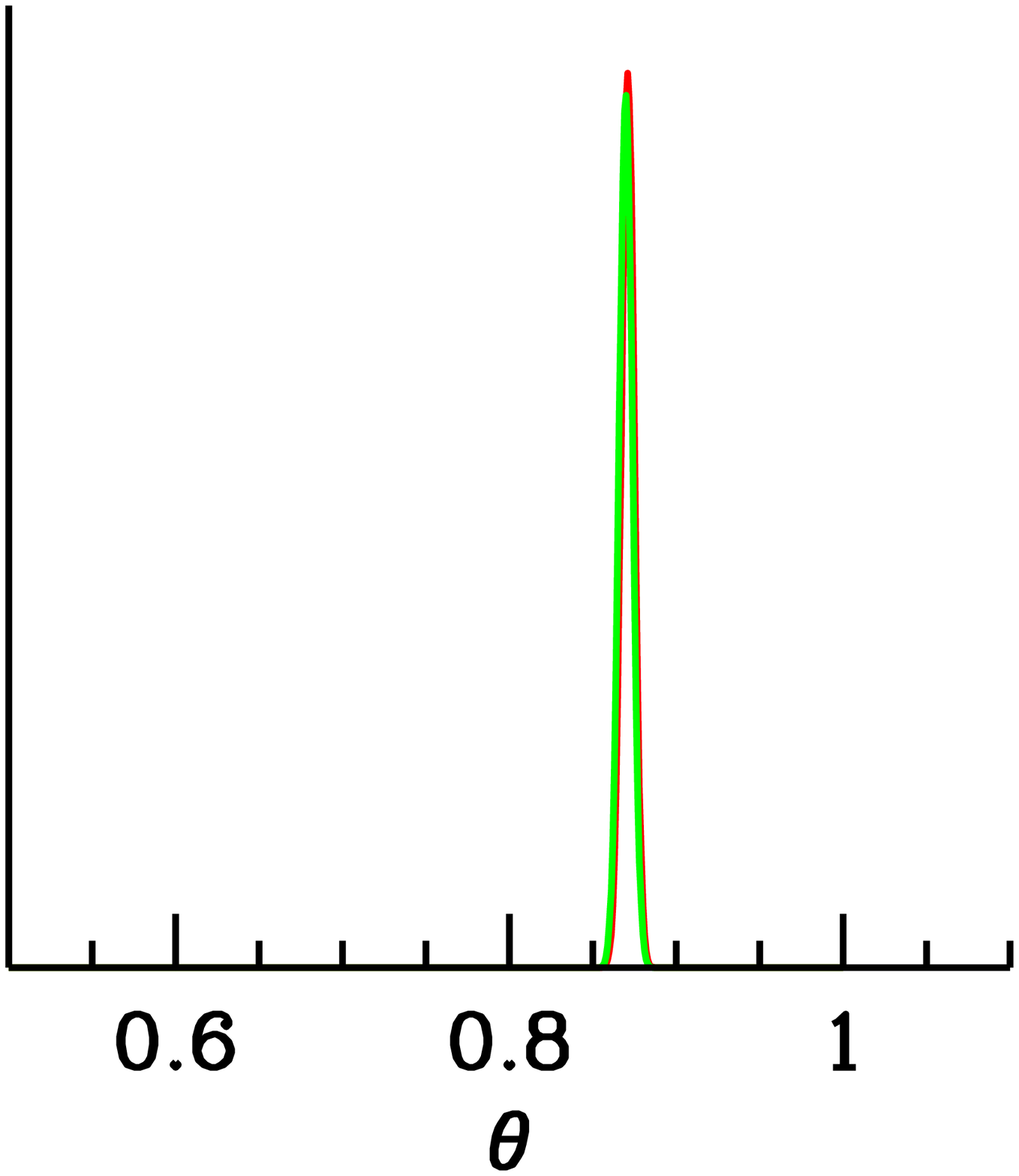}\hfill
\caption{Converging views in Bayesian inference. Two scientists
having different prior believes $p(\params|I_i)$ about the value
of a quantity $\theta$ (panel (a), red and green pdf's) observe
one datum with likelihood $\like(\params)$ (panel (b)), after
which their posteriors $p(\theta|m_1)$ (panel (c), obtained via
Bayes Theorem, Eq.~\eqref{eq:Bayes_Theorem_simple}) represent
their updated states of knowledge on the parameter. After
observing 100 data points, the two posteriors have become
essentially indistinguishable (d). }
\label{fig:prior_to_posterior}
\end{figure}

Finally, objectivity is ensured by the fact that two scientists in
the same state of knowledge should assign the same prior, hence
their posterior are identical if they observe the same data. The
fact that the prior assignment eventually becomes irrelevant as
better and better data make the posterior likelihood--dominated in
uncontroversial in principle but may be problematic in practice.
Often the data are not strong enough to override the prior, in
which case great care must be given in assessing how much of the
final inference depends on the prior choice. This might occur for
small sample sizes, or for problems where the dimensionality of
the hypothesis space is larger than the number of observations
(for example, in image reconstruction). Even in such a case, if
different prior choices lead to different posteriors we can still
conclude that the data are not informative enough to completely
override our prior state of knowledge and hence we have learned
something useful about the constraining power (or lack thereof) of
the data.

The situation is somewhat different for model comparison
questions. In this case, it is precisely the available prior
volume that is important in determining the penalty that more
complex models with more free parameters should incur into (this
is discussed in detail in section~\ref{sec:modcomp}). Hence the
impact of the prior choice is much stronger when dealing with
model selection issues, and care should be exercised in assessing
how much the outcome would change for physically reasonable
changes in the prior.

There is a vast literature on priors that we cannot begin to
summarize here. Important issues concern the determination of
``ignorance priors'', i.e. priors reflecting a state of
indifference with respect to symmetries of the problem considered.
``Reference priors'' exploit the idea of using characteristics of
what the experiment is expected to provide to construct priors
representing the least informative state of knowledge. In order to
be probability distributions, priors must be {\em proper}, i.e.
normalizable to unity probability content. ``Flat priors'' are
often a standard choice, in which the prior is taken to be
constant within some minimum and maximum value of the parameters,
i.e. for a 1--dimensional case $p(\params) = (\params_{\rm max} -
\params_{\rm min})^{-1}$. The rationale is that we should assign
equal probability to equal states of knowledge. However, flat
priors are not always as harmless as they appear. One reason is
that a flat prior on a parameter $\params$ does not correspond to
a flat prior on a non--linear function of that parameter,
$\psi(\params)$. The two priors are related by
 \begin{equation}
 \label{eq:prior_transformation}
 p(\psi) = p(\params) {\Big\arrowvert} \frac{\dr \params}{\dr \psi} {\Big\arrowvert},
 \end{equation}
so for a non--linear dependence $\psi(\params)$ the term $|\dr
\params/\dr \psi|$ means that an uninformative (flat)
prior on $\params$ might be strongly informative about $\psi$ (in
the multi--dimensional case, the derivative term is replaced by
the determinant of the Jacobian for the transformation).
Furthermore, if we are ignorant about the {\em scale} of a
quantity $\params$, it can be shown (see e.g.~\cite{JaynesBook})
that the appropriate prior is flat on $\ln \params$, which gives
equal weight to all orders of magnitude. This prior corresponds to
$p(\params) \propto
\params^{-1}$ and is called ``Jeffreys' prior''. It is appropriate
for example for the rate of a Poisson distribution. For further
details on prior choice, and especially so--called ``objective
priors'', see e.g.~chapter 5 in~\cite{Ghosh:2006} and references
therein.

\section{Bayesian parameter inference}
\label{sec:Bayesian_parameter_inference}

\subsection{The general problem and its solution}
\label{sec:inference_theory}

The general problem of Bayesian parameter inference can be
specified as follows. We first choose a model containing a set of
hypotheses in the form of a vector of parameters, $\params$. The
parameters might describe any aspect of the model, but usually
they will represent some physically meaningful quantity, such as
for example the mass of an extra--solar planet or the abundance of
dark matter in the Universe. Together with the model we must
specify the priors for the parameters. Priors should summarize our
state of knowledge about the parameters before we consider the new
data, and for the parameter inference step the prior for a new
observation might be taken to be the posterior from a previous
measurement (for model comparison issues the prior is better
understood in a different way, see section~\ref{sec:modcomp}). The
{\em caveats} about priors and prior specifications presented in
the previous section will apply at this stage.

The central step is to construct the likelihood function for the
measurement, which usually reflects the way the data are obtained.
For example, a measurement with Gaussian noise will be represented
by a Normal distribution, while $\gamma$--ray counts on a detector
will have a Poisson distribution for a likelihood. Nuisance
parameters related to the measurement process might be present in
the likelihood, e.g. the variance of the Gaussian might be unknown
or the background rate in the absence of the source might be
subject to uncertainty. This is no matter of concern for a
Bayesian, as the general strategy is always to work out the joint
posterior for all of the parameters in the problem and then
marginalize over the ones we are not interested in. Assuming that
we have a set of physically interesting parameters $\phi$ and a
set of nuisance parameters $\psi$, the joint posterior for
$\params = (\phi, \psi)$ is obtained through Bayes' Theorem:
 \begin{equation}\label{eq:Bayes_Theorem}
 p(\params | \data, \mdl) = \like(\params) \frac{p(\params | \mdl)}{p(\data | \mdl)}
 \end{equation}
where we have made explicit the choice of a model $\mdl$ by
writing it on the righ--hand--side of the conditioning symbol.
Recall that $\like(\params ) \equiv p(\data | \params, \mdl)$
denotes the likelihood and $p(\params| \mdl)$ the prior. The
normalizing constant $p(\data | \mdl)$ (``the Bayesian evidence'')
is irrelevant for parameter inference (but central to model
comparison, see section~\ref{sec:modcomp}), so we can write the
marginal posterior on the parameter of interest as (marginalizing
over the nuisance parameters)
 \begin{equation} \label{eq:marginal_posterior}
 p(\phi| \data, \mdl) \propto \int \like(\phi, \psi) p(\phi, \psi |
 \mdl)\dr \psi.
 \end{equation}
The final inference on $\phi$ from the posterior can then be
communicated either by some summary statistics (such as the mean,
the median or the mode of the distribution, its standard deviation
and the correlation matrix among the components) or more usefully
(especially for cases where the posterior presents multiple peaks
or heavy tails) by plotting one or two dimensional subsets of
$\phi$, with the other components marginalized over.

In real life there are only a few cases of interest for which the
above procedure can be carried out analytically. Quite often,
however, the simple case of a Gaussian prior and a Gaussian
likelihood can offer useful guidance regarding the behaviour of
more complex problems. \ct{An analytical model of a
Poisson--distributed likelihood for estimating source counts in
the presence of a background signal is worked out
in~\cite{Loredo:1990}.} In general, however, actual problems in
cosmology and astrophysics are not analytically tractable and one
must resort to numerical techniques to evaluate the likelihood and
to draw samples from the posterior. Fortunately this is not a
major hurdle thanks to the recent increase of cheap computational
power. In particular, numerical inference often employs a
technique called {\em Markov Chain Monte Carlo}, which allows to
map out numerically the posterior distribution of
Eq.~\eqref{eq:Bayes_Theorem} even in the most complicated
situations, where the likelihood can only be obtained by numerical
simulation, the parameter space can have hundreds of dimensions
and the posterior has multiple peaks and a complicated structure.

\ct{For further reading about Bayesian parameter inference,
see~\cite{Babu:1996,Sivia:1996,Gelman:2003,Gregory:2005}. For more
advanced applications to problems in astrophysics and cosmology,
see~\cite{Feigelson:2003,Liddle:BMIC2008}.}

\subsection{Markov Chain Monte Carlo techniques for parameter inference}
\label{sec:MCMC}

The general solution to any inference problem has been outlined in
the section above: it remains to find a way to evaluate the
posterior of Eq.~\eqref{eq:Bayes_Theorem} for the usual case where
analytical solutions do not exist or are insufficiently accurate.
Nowadays, Bayesian inference heavily relies on numerical
simulation, in particular in the form of Markov Chain Monte Carlo
(MCMC) techniques, which are discussed in this section.

The purpose of the Markov chain Monte Carlo algorithm is to
construct a sequence of points in parameter space (called ``a
chain''), whose density is proportional to the posterior pdf of
Eq.~\eqref{eq:Bayes_Theorem}. Developing a full theory of Markov
chains is beyond the scope of the present article\ct{ (see
e.g.~\cite{Robert:2004,Gamerman:2006} instead)}. For our purposes
it suffices to say that a Markov chain is defined as a sequence of
random variables $\{X^{(0)}, X^{(1)}, \dots, X^{(M-1)}\}$ such
that the probability of the $(t+1)$--th element in the chain only
depends on the value of the $t$--th element. The crucial property
of Markov chains is that they can be shown to converge to a
stationary state (i.e., which does not change with $t$) where
successive elements of the chain are samples from the {\em target
distribution}, in our case the posterior $p(\params|\data)$. The
generation of the elements of the chain is probabilistic in
nature, and several algorithms are available to construct Markov
chains. The choice of algorithm is highly dependent on the
characteristics of the problem at hand, and ``tayloring'' the MCMC
to the posterior one wants to explore often takes a lot of effort.
Popular and effective algorithms include the Metropolis--Hastings
algorithm\ct{~\cite{Metropolis:1953am,Hastings:1970}}, Gibbs
sampling\ct{~(see e.g.~\cite{Smith:1993})}, Hamiltonian Monte
Carlo\ct{~(see e.g.~\cite{Hanson:2001}} and importance sampling.

Once a Markov chain has been constructed, obtaining Monte Carlo
estimates of expectations for any function of the parameters
becomes a trivial task. For example, the posterior mean is given
by (where $\langle \cdot \rangle$ denotes the expectation value
with respect to the posterior)
 \begin{equation} \label{eq:expectation}
 \langle \params \rangle \approx \int  p(\params|\data)\params \dr\params
   = \frac{1}{M} \sum_{t=0}^{M-1} \params^{(t)},
 \end{equation}
where the equality with the mean of the samples from the MCMC
follows because the samples $\params^{(t)}$ are generated from the
posterior by construction. In general, one can easily obtain the
expectation value of any function of the parameters $f(\params)$
as
 \begin{equation} \label{eq:MC_estimate}
 \langle f(\params) \rangle \approx \frac{1}{M}\sum_{t=0}^{M-1}  f(\params^{(t)}).
 \end{equation}
It is usually interesting to summarize the results of the
inference by giving the 1--dimensional {\em marginal probability}
for the $j$--th element of $\params$, $\params_j$. Taking without
loss of generality $j=1$ and a parameter space of dimensionality
$n$, the equivalent expression to
Eq.~\eqref{eq:def_marginalisation} for the case of continuous
variables is
 \begin{equation} \label{eq:marginalisation_continuous}
 p(\theta_1|\data) = \int  p(\params|\data) \dr \theta_2 \dots \dr
 \theta_n,
 \end{equation}
where $p(\theta_1|\data)$ is the {\em marginal posterior} for the
parameter $\theta_1$. From the Markov chain it is trivial to
obtain and plot the marginal posterior on the left--hand--side of
Eq.~\eqref{eq:marginalisation_continuous}: since the elements of
the Markov chains are samples from the full posterior,
$p(\params|\data)$, their density reflects the value of the full
posterior pdf. It is then sufficient to divide the range of
$\theta_1$ in a series of bins and {\em count the number of
samples falling within each bin}, simply ignoring the coordinates
values $\theta_2, \dots, \theta_n$. A 2--dimensional posterior is
defined in an analogous fashion.

There are several important practical issues in working with MCMC
methods\ct{~(for details see e.g.~\cite{Robert:2004})}. Especially
for high--dimensional parameter spaces with multi--modal
posteriors it is important {\em not} to use MCMC techniques as a
black box, since poor exploration of the posterior can lead to
serious mistakes in the final inference if it remains undetected.
Considerable care is required to ensure as much as possible that
the MCMC exploration has covered the relevant parameter space.

\section{Bayesian model comparison}
\label{sec:modcomp}

\subsection{Shaving theories with Occam's razor}

When there are several competing theoretical models, Bayesian
model comparison provides a formal way of evaluating their
relative probabilities in light of the data and any prior
information available. The ``best'' model is then the one which
strikes an optimum balance between quality of fit and
predictivity. In fact, it is obvious that a model with more free
parameters will always fit the data better (or at least as good
as) a model with less parameters. However, more free parameters
also mean a more ``complex'' model, in a sense that we will
quantify below in section~\ref{sec:complexity}. Such an added
complexity ought to be avoided whenever a simpler model provides
an adequate description of the observations. This guiding
principle of simplicity and economy of an explanation is known as
{\em Occam's razor} --- the simplest theory compatible with the
available evidence ought to be preferred\footnote{In its
formulation by the medieval English philosopher and Franciscan
monk William of Ockham (ca. 1285-1349): ``Pluralitas non est
ponenda sine neccesitate''.}. Bayesian model comparison offers a
formal way to evaluate whether the extra complexity of a model is
required by the data, thus putting on a firmer statistical grounds
the evaluation and selection process of scientific theories that
scientists often carry out at a more intuitive level. For example,
a Bayesian model comparison of the Ptolemaic model of epicycles
versus the heliocentric model based on Newtonian gravity would
favour the latter because of its simplicity and ability to explain
planetary motions in a more economic fashion than the baroque
construction of epicycles.

An important feature is that an alternative model must be
specified against which the comparison is made. In contrast with
frequentist goodness--of--fit tests (such as chi--square tests),
Bayesian model comparison maintains that it is pointless to reject
a theory unless an alternative explanation is available that fits
the observed facts better\ct{ (for more details about the
difference in approach with frequentist hypothesis testing,
see~\cite{Loredo:1990})}. In other words, unless the observations
are totally impossible within a model, finding that the data are
improbable given a theory does not say anything about the
probability of the theory itself {\em unless we can compare it
with an alternative}. A consequence of this is that the
probability of a theory that makes a correct prediction can
increase if the prediction is confirmed by observations, provided
competitor theories do not make the same prediction. This agrees
with our intuition that a verified prediction lends support to the
theory that made it, in contrast with the limited concept of
falsifiability advocated by Popper (i.e., that scientific theories
can only be tested by proving them wrong). So for example,
perturbations to the motion of Uranus led the French astronomer
U.J.J Leverrier and the English scholar J.C. Adams to formulate
the prediction, based on Newtonian theory, that a further planet
ought to exist beyond the orbit of Uranus. The discovery of
Neptune in 1846 within 1 degree of the predicted position thus
should strengthen our belief in the correctness of Newtonian
gravity. However, as discussed in detail in chapter 5 of
Ref~\cite{JaynesBook}, the change in the plausibility of Newton's
theory following the discovery of Uranus crucially depends on the
alternative we are considering. If the alternative theory is
Einstein gravity, then obviously the two theories make the same
predictions as far as the orbit of Uranus is concerned, hence
their relative plausibility is unchanged by the discovery. The
alternative ``Newton theory is false'' is not useful in Bayesian
model comparison, and we are forced to put on the table a more
specific model than that before we can assess how much the new
observation changes our relative degree of belief between an
alternative theory and Newtonian gravity.

In the context of model comparison it is appropriate to think of a
model as a specification of a set of parameters $\params$ {\em
and} of their prior distribution, $p(\params | \mdl$). As shown
below, it is the number of free parameters {\em and} their prior
range that control the strength of the Occam's razor effect in
Bayesian model comparison: models that have many parameters that
can take on a wide range of values but that are not needed in the
light of the data are penalized for their unwarranted complexity.
Therefore, {\em the prior choice ought to reflect the available
parameter space under the model $\mdl$, independently of
experimental constraints we might already be aware of}. This is
because we are trying to assess the economy (or simplicity) of the
model itself, and hence the prior should be based on theoretical
or physical constraints on the model under consideration. Often
these will take the form of a range of values that are deemed
``intuitively'' plausible, or ``natural''. Thus the prior
specification is inherent in the model comparison approach.

The prime tool for model selection is the {\em Bayesian evidence},
discussed in the next three sections.  A quantitative measure of
the {\em effective model complexity} is introduced  in
section~\ref{sec:complexity}. We then present some popular
approximations to the full Bayesian evidence that attempt to avoid
the difficulty of priors choice, the {\em information criteria},
and discuss the limits of their applicability in
section~\ref{sec:information_criteria}.

\ct{For reviews on model selection see
e.g.~\cite{Jeffreys:1961,Burnham:2002,MacKay:2003}
and~\cite{Mukherjee:2008} for cosmological applications. Good
starting points on Bayes factors are~\cite{Kass:1995,Han:2001}. A
discussion of the spirit of model selection can be found in the
first part of~\cite{Garrett:1993}.}

\subsection{The Bayesian evidence}

The evaluation of a model's performance in the light of the data
is based on the {\em Bayesian evidence}, which in the statistical
literature is often called {\em marginal likelihood} or {\em model
likelihood}. Here we follow the practice of the cosmology and
astrophysics community and will use the term ``evidence'' instead.
The evidence is the normalization integral on the
right--hand--side of Bayes' theorem, Eq.~\eqref{eq:evidence_def},
which we rewrite here for a continuous parameter space
$\Omega_\mdl$ and conditioning explicitly on the model under
consideration, $\mdl$:
 \be
\label{eq:evidence_def_mdl}
 p(\data | \mdl) \equiv {\int_{\Omega_\mdl} p(\data | \params, \mdl)
p(\params | \mdl)\dr\params} \quad({\rm Bayesian~evidence }).
 \ee
Thus the Bayesian evidence is the average of the likelihood under
the prior for a specific model choice. From the evidence, the
model posterior probability given the data is obtained by using
Bayes' Theorem to invert the order of conditioning:
 \be
 p(\mdl|\data) \propto p(\mdl)p(\data|\mdl),
 \ee
where we have dropped an irrelevant normalization constant that
depends only on the data and $p(\mdl)$ is the prior probability
assigned to the model itself. Usually this is taken to be
non--committal and equal to $1/N_m$ if one considers $N_m$
different models. When comparing two models, $\mdl_0$ versus
$\mdl_1$, one is interested in the ratio of the posterior
probabilities, or {\em posterior odds}, given by
 \be \label{eq:posterior_odds}
 \frac{p(\mdl_0|\data)}{p(\mdl_1|\data)} = B_{01}
 \frac{p(\mdl_0)}{p(\mdl_1)}
\ee
 and the {\em Bayes factor} $B_{01}$ is the ratio of the
models' evidences:
\begin{equation}
 \label{eq:Bayes_factor}
 B_{01} \equiv \frac{p(\data | \mdl_0)}{p(\data | \mdl_1)} \quad({\rm Bayes~factor}).
\end{equation}
A value $B_{01} > (<)~1$ represents an increase (decrease) of the
support in favour of model 0 versus model 1 given the observed
data. From Eq.~\eqref{eq:posterior_odds} it follows that the Bayes
factor gives the factor by which the relative odds between the two
models have changed after the arrival of the data, regardless of
what we thought of the relative plausibility of the models before
the data, given by the ratio of the prior models' probabilities.
Therefore the relevant quantity to update our state of belief in
two competing models is the Bayes factor.

To gain some intuition about how the Bayes factor works, consider
two competing models: $\mdl_0$ predicting that a quantity $\theta
= 0$ with no free parameters, and $\mdl_1$ which assigns $\theta$
a Gaussian prior distribution with 0 mean and variance $\Sigma^2$.
Assume we perform a measurement of $\theta$ described by a normal
likelihood of standard deviation $\sigma$, and with the maximum
likelihood value lying $\lambda$ standard deviations away from 0,
i.e. $|\pml/\sigma| = \lambda$. Then the Bayes factor between the
two models is given by, from Eq.~\eqref{eq:Bayes_factor}
 \be \label{eq:B01_example}
 B_{01} = \sqrt{1 + (\sigma/\Sigma)^{-2}} \exp \left( -
 \frac{\lambda^2}{2(1 + (\sigma/\Sigma)^2)} \right).
 \ee
For $\lambda \gg 1$, corresponding to a detection of the new
parameter at many sigma, the exponential term dominates and
$B_{01} \ll 1$, favouring the more complex model with a non--zero
extra parameter, in agreement with the usual conclusion. But if
$\lambda \lsim 1$ and $\sigma/\Sigma \ll 1$ (i.e., the likelihood
is much more sharply peaked than the prior and in the vicinity of
0), then the prediction of the simpler model that $\theta = 0$ has
been confirmed. This leads to the Bayes factor being dominated by
the Occam's razor term, and $B_{01} \approx \Sigma/\sigma$, i.e.
evidence accumulates in favour of the simpler model proportionally
to the volume of ``wasted'' parameter space. If however
$\sigma/\Sigma \gg 1$ then the likelihood is less informative than
the prior and $B_{01} \rightarrow 1$, i.e. the data have not
changed our relative belief in the two models.

Bayes factors are usually interpreted against the Jeffreys'
scale~\cite{Jeffreys:1961} for the strength of evidence, given in
Table~\ref{Tab:Jeff}. This is an empirically calibrated scale,
with thresholds at values of the odds of about $3:1$, $12:1$ and
$150:1$, representing weak, moderate and strong evidence,
respectively. A useful way of thinking of the Jeffreys' scale is
in terms of betting odds --- many of us would feel that odds of
$150:1$ are a fairly strong disincentive towards betting a large
sum of money on the outcome. Also notice from Table~\ref{Tab:Jeff}
that the relevant quantity in the scale is the logarithm of the
Bayes factor, which tells us that evidence only accumulates slowly
and that indeed moving up a level in the evidence strength scale
requires about an order of magnitude more support than the level
before.

\begin{table}
\tbl{Empirical scale for evaluating the strength of evidence when
comparing two models, $\mdl_0$ versus $\mdl_1$ (so--called
``Jeffreys' scale''). Threshold values are empirically set, and
they occur for values of the logarithm of the Bayes factor of
$|\ln B_{01}|=1.0$, 2.5 and 5.0. The right--most column gives our
convention for denoting the different levels of evidence above
these thresholds. The probability column refers to the posterior
probability of the favoured model, assuming non--committal priors
on the two competing models, \ie~$p(\mdl_0) = p(\mdl_1) = 1/2$ and
that the two models exhaust the model space, $p(\mdl_0|\data) +
p(\mdl_1|\data) = 1$.\label{Tab:Jeff} }
 {\begin{tabular}{l l l l} \toprule
  $|\ln B_{01}|$ & Odds & Probability & Strength of evidence \\\colrule
 $<1.0$ & $\lsim 3:1$ & $<0.750$ & Inconclusive \\
 $1.0$ & $\sim 3:1$ & $0.750$ & Weak evidence \\
 $2.5$ & $\sim 12:1$ & $0.923$ & Moderate evidence \\
 $5.0$ & $\sim 150:1$ & $0.993$ & Strong evidence \\
 \botrule
\end{tabular}}
\end{table}

Bayesian model comparison {\em does not} replace the parameter
inference step (which is performed within each of the models
separately). Instead, model comparison {\em extends} the
assessment of hypotheses in the light of the available data to the
space of theoretical models, as evident from
Eq.~\eqref{eq:posterior_odds}, which is the equivalent expression
for models to Eq.~\eqref{eq:Bayes_Theorem}, representing inference
about the parameters value within each model (for multi--model
inference, merging the two levels, see
section~\ref{sec:other_uses_evidence}).

\subsection{Computation and interpretation of the evidence}

The computation of the Bayesian
evidence~\eqref{eq:evidence_def_mdl} is in general a numerically
challenging task, as it involves a multi--dimensional integration
over the whole of parameter space. An added difficulty is that the
likelihood is often sharply peaked within the prior range, but
possibly with long tails that do contribute significantly to the
integral and which cannot be neglected. Other problematic
situations arise when the likelihood is multi--modal, or when it
has strong degeneracies that confine the posterior to thin sheets
in parameter space. Until recently, the application of Bayesian
model comparison has been hampered by the difficulty of reliably
estimating the evidence. Fortunately, several methods are now
available, each with its own strengths and domains of
applicability.

\begin{enumerate}

\item The numerical method of choice until recently has been
thermodynamic integration, also
 called {\em simulated annealing} (see e.g.
\xcite{Gregory:2005,Press:2007}{Clyde:2006} and references therein
for details). Its computational cost can become fairly large, as
it depends heavily on the dimensionality of the parameter space
and on the characteristic of the likelihood function. In typical
cosmological
applications~\cite{Slosar:2002dc,Beltran:2005xd,Bridges:2005br},
thermodynamic integration can require up to $10^7$ likelihood
evaluations, two orders of magnitude more than MCMC--based
parameter estimation.

\item Skilling~\cite{SkillingNS,Skilling:2006} has put forward an
elegant algorithm called ``nested sampling'', which has been
implemented in the cosmological context
by~\cite{Bassett:2004wz,Mukherjee:2005wg,Shaw:2007jj,Feroz:2007kg,Bridges:2006mt}\ct{
(for a theoretical discussion of the algorithmic properties,
see~\cite{Chopin:2007})}. The gist of nested sampling is that the
multi--dimensional evidence integral is recast into a
one--dimensional integral that is easy to evaluate numerically.
This technique allows to reduce the computational burden to about
$10^5$ likelihood evaluations\footnote{Publicly available modules
implementing nested sampling can be found at
\texttt{cosmonest.org}~\cite{Mukherjee:2005wg} and
\texttt{http://www.mrao.cam.ac.uk/software/cosmoclust/}~\cite{Shaw:2007jj}
(accessed Feb 2008).}. Recently, the development of what is called
``multi--modal nested sampling'' has allowed to increase
significantly the efficiency of the method~\cite{Feroz:2007kg},
reducing the number of likelihood evaluations by another order of
magnitude.

 \item Approximations to the Bayes factor, Eq.~\eqref{eq:Bayes_factor},
are available for situations in which the models being compared
are {\em nested} into each other, i.e. the more complex model
($\mdl_1$) reduces to the original model ($\mdl_0$) for specific
values of the new parameters. This is a fairly common scenario in
cosmology, where one wishes to evaluate whether the inclusion of
the new parameters is supported by the data. For example, we might
want to assess whether we need isocurvature contributions to the
initial conditions for cosmological perturbations, or whether a
curvature term in Einstein's equation is needed, or whether a
non--scale invariant distribution of the primordial fluctuation is
preferred (see Table~\ref{tab:evidence_summary} for actual
results). Writing for the extended model parameters $\params =
(\pzero, \pone)$, where the simpler model $\mdl_0$ is obtained by
setting $\pone = 0$, and assuming further that the prior is
separable (which is usually the case in cosmology), i.e.\ that
 \begin{equation}
 p(\pzero, \pone| \mdl_1) = p(\pone | \mdl_1) p(\pzero | \mdl_0),
 \end{equation}
the Bayes factor can be written in all generality as
 \begin{equation} \label{eq:savagedickey}
 B_{01} = \left.\frac{p(\pone \vert \data, \mdl_1)}{p(\pone |
 \mdl_1)}\right|_{\pone = 0}.
 \end{equation}
This expression is known as the Savage--Dickey density ratio
(SDDR, see \cite{Verdinelli:1995} and references therein). The
numerator is simply the marginal posterior under the more complex
model evaluated at the simpler model's parameter value, while the
denominator is the prior density of the more complex model
evaluated at the same point. This technique is particularly useful
when testing for one extra parameter at the time, because then the
marginal posterior $p(\pone \vert \data, \mdl_1)$ is a
1--dimensional function and normalizing it to unity probability
content only requires a 1--dimensional integral, which is simple
to do using for example the trapezoidal rule.

\item An instructive approximation to the Bayesian evidence can be
obtained when the likelihood function is unimodal and
approximately Gaussian in the parameters. Expanding the likelihood
around its peak to second order one obtains the Laplace
approximation
 \be \label{eq:like_Gaussian}
 p(\data | \params, \mdl) \approx
 \lmax
 \exp\left[-\frac{1}{2}(\params-\pml)^t L
(\params-\pml)\right],
 \ee
where $\pml$ is the maximum--likelihood point, $\lmax$ the maximum
likelihood value and $L$ the likelihood Fisher matrix (which is
the inverse of the covariance matrix for the parameters). Assuming
as a prior a multinormal Gaussian distribution with zero mean and
Fisher information matrix $P$ one obtains for the evidence,
Eq.~\eqref{eq:evidence_def_mdl}
 \be
 \label{eq:evidence_Gaussian}
 p(\data | \mdl) = \lmax \frac{|F|^{-1/2}}{|P|^{-1/2}}
 \exp\left[-\frac{1}{2}({\pml}^t L \pml - \overline{\params}^tF\overline{\params})
 \right],
 \ee
where the posterior Fisher matrix is $F = L + P$ and the posterior
mean is given by $\overline{\params} = F^{-1}L \pml$.
\end{enumerate}

From Eq.~\eqref{eq:evidence_Gaussian} we can deduce a few
qualitatively relevant properties of the evidence. First, the
quality of fit of the model is expressed by $\lmax$, the best--fit
likelihood. Thus a model which fits the data better will be
favoured by this term. The term involving the determinants of $P$
and $F$ is a volume factor, encoding the Occam's razor effect. As
$|P|\leq|F|$, it penalizes models with a large volume of wasted
parameter space, \ie\ those for which the parameter space volume
$|F|^{-1/2}$ which survives after arrival of the data is much
smaller than the initially available parameter space under the
model prior, $|P|^{-1/2}$. Finally, the exponential term
suppresses the likelihood of models for which the parameters
values which maximise the likelihood, $\pml$, differ appreciably
from the expectation value under the posterior,
$\overline{\params}$. Therefore when we consider a model with an
increased number of parameters we see that {\em its evidence will
be larger only if the quality--of--fit increases enough to offset
the penalizing effect of the Occam's factor} (see also the
discussion in~\cite{Heavens:2007ka}).

On the other hand, it is important to notice that the Bayesian
evidence does {\em not} penalizes models with parameters that are
unconstrained by the data. It is easy to see that unmeasured
parameters (i.e., parameters whose posterior is equal to the
prior) do not contribute to the evidence integral, and hence model
comparison does not act against them, awaiting better data.

\subsection{The rough guide to model comparison}

The gist of Bayesian model comparison can be summarized by the
following, back--of--the--envelope Bayes factor computation for
nested models. The result is surprisingly close to what one would
obtain from the more elaborate, fully--fledged evidence
evaluation, and can serve as a rough guide for the Bayes factor
determination.

\begin{figure}
\centerline{\includegraphics[width=0.5\linewidth]{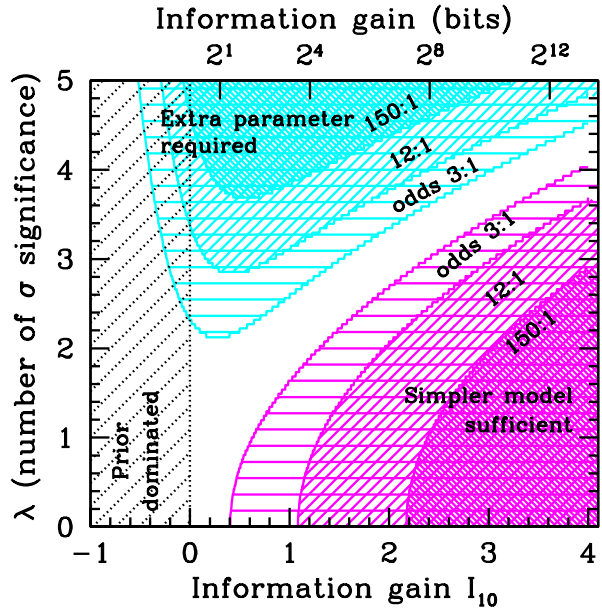}}
\caption{Illustration of Bayesian model comparison for two nested
models, where the more complex model has one extra parameter. The
outcome of the model comparison depends both on the information
content of the data with respect to the {\em a priori} available
parameter space, $\It$ (horizontal axis) and on the quality of fit
(vertical axis, $\lambda$, which gives the number of sigma
significance of the measurement for the extra parameter). The
contours are computed from Eq.~\eqref{eq:savagedickey}, assuming a
Gaussian likelihood and prior (adapted
from~\cite{Trotta:2005ar}).} \label{fig:evidence_plan}
\end{figure}

Returning to the example of Eq.~\eqref{eq:B01_example}, if the
data are informative with respect to the prior on the extra
parameter (i.e., for $\sigma/\Sigma \ll 1$) the logarithm of the
Bayes factor is given approximately by
 \begin{equation}
 \label{eq:B01_info}
 \ln B_{01} \approx \ln \left( \Sigma/\sigma \right) -
 \lambda^2/2,
 \end{equation}
where as before $\lambda$ gives the number of sigma away from a
null result (the ``significance'' of the measurement). The first
term on the right--hand--side is approximately the logarithm of
the ratio of the prior to posterior volume. We can interpret it as
the information content of the data, as it gives the factor by
which the parameter space has been reduced in going from the prior
to the posterior. This term is positive for informative data, i.e.
if the likelihood is more sharply peaked than the prior. The
second term is always negative, and it favours the more complex
model if the measurement gives a result many sigma away from the
prediction of the simpler model (i.e., for $\lambda \gg 0$). We
are free to measure the information content in base--10 logarithm
(as this quantity is closer to our intuition, being the order of
magnitude of our information increase), and we define the quantity
$\It \equiv \log_{10}\left(\Sigma/\sigma \right)$.
Figure~\ref{fig:evidence_plan} shows contours of $\vert \ln B_{01}
\vert = $~const for const~$= 1.0,2.5,5.0$ in the $(\It, \lambda)$
plane, as computed from Eq.~\eqref{eq:B01_info}. The contours
delimit significative levels for the strength of evidence,
according to the Jeffreys' scale (Table~\ref{Tab:Jeff}). For
moderately informative data ($\It \approx 1 - 2$) the measured
mean has to lie at least about $4\sigma$ away from 0 in order to
robustly disfavor the simpler model (i.e., $\lambda \gsim 4$).
Conversely, for $\lambda \lsim 3$ highly informative data ($\It
\gsim 2$) do favor the conclusion that the extra parameter is
indeed 0. In general, a large information content favors the
simpler model, because Occam's razor penalizes the large volume of
``wasted'' parameter space of the extended model.

An useful properties of Figure~\ref{fig:evidence_plan} is that the
impact of a change of prior can be easily quantified. A different
choice of prior width (i.e., $\Sigma$) amounts to a {\em
horizontal shift} across Figure~\ref{fig:evidence_plan}, at least
as long as $\It>0$ (i.e., the posterior is dominated by the
likelihood). Picking more restrictive priors (reflecting more
predictive theoretical models) corresponds to shifting the result
of the model comparison to the left of
Figure~\ref{fig:evidence_plan}, returning an inconclusive result
(white region) or a prior--dominated outcome (hatched region).
Notice that results in the 2--3 sigma range, which are fairly
typical in cosmology, can only support the more complex model in a
very mild way at best (odds of $3:1$ at best), while actually
being most of the time either inconclusive or in favour of the
simpler hypothesis (pink shaded region in the bottom right
corner).

Bayesian model comparison is usually {\em conservative} when it
comes to admitting a new quantity in our model, even in the case
when the prior width is chosen incorrectly. Consider the following
two possibilities:
 \begin{itemize}
 \item  If the prior range is too small, the model
comparison result will be non--committal (white region in
Figure~\ref{fig:evidence_plan}), or even prior dominated (hatched
region, where the posterior is dominated by the prior). Hence in
this case we have to hold judgement until better data come along.
 \item Too wide a prior will instead unduly favour the simpler model (pink, shaded
regions). However, as new, better data come along the result will
move to the right (for a fixed prior width, as the likelihood
becomes narrower) but eventually also upwards, towards a larger
number of sigma significance, if the true model really has a
non--zero extra parameter. Eventually, our initial ``poor'' prior
choice will be overridden as the number of sigma becomes large
enough to take the result into the blue, shaded region.
 \end{itemize}

In both cases the result of the model comparison will eventually
override the ``wrong'' prior choice (although it might take a long
time to do so), exactly as it happens for parameter inference.

\subsection{Getting around the prior -- The maximal evidence for a new parameter}
\label{sec:Bayesian_p_vals}

For nested models, Eq.~\eqref{eq:savagedickey} shows that the
relative probability of the more complex model can be made
arbitrarily small by increasing the broadness of the prior for the
extra parameters, $p(\pone | \mdl_1)$ (as the prior is a pdf, it
must integrate to unit probability. Hence a broader prior
corresponds to a smaller value of $p(\pone | \mdl_1)_{\pone = 0}$
in the denominator). Often, this is not problematical as prior
ranges for the new parameters can (and should) be motivated from
the underlying theory. For example, in assessing whether the
scalar spectral index ($n_s$) of the primordial perturbations
differs from the scale--invariant value $n_s=1$, the prior range
of the index can be constrained to be $0.8\lsim n_s \lsim 1.2$
within the theoretical framework of slow roll inflation (more on
this in section~\ref{sec:cosmo_model_building}). The sensitivity
of the model comparison result can also be investigated for other
plausible, physically motivated choices of prior ranges, see
e.g.~\cite{Trotta:2007hy,Trotta:2006ww}. If the model comparison
outcome is qualitatively the same for a broad choice of plausible
priors, then we can be confident that the result is robust.

Although the Bayesian evidence offers a well--defined framework
for model comparison, there are cases where there is not a
specific enough model available to place meaningful limits on the
prior ranges of new parameters in a model. This hurdle arises
frequently in cases when the new parameters are a phenomenological
description of a new effect, only loosely tied to the underlying
physics, such as for example expansion coefficients of some
series. An interesting possibility in such a case is to choose the
prior on the new parameters in such a way as to {\em maximise the
probability of the new model}, given the data. If, even under this
best case scenario, the more complex model is not significantly
more probable than the simpler model, then one can confidently say
that the data does not support the addition of the new parameters,
without worrying that some other choice of prior will make the new
model more
probable~\xcite{Sellke:2001}{Berger:1987a,Berger:1987b}.

Consider the Bayes factor in favour of the more complex model,
$B_{10} \equiv 1/B_{01}$, with $B_{01}$ given by
Eq.~\eqref{eq:Bayes_factor}. The simpler model, $\mdl_0$, is
obtained from $\mdl_0$ by setting $\params = \params^*$. An
absolute upper bound to the evidence in favour of the more complex
model is obtained by choosing $p(\params|\mdl_1)$ to be a delta
function centered at the maximum likelihood value under $\mdl_1$,
$\pml$. It can be shown that in this case the upper bound $\bB$
corresponds to the likelihood ratio between $\pml$ and
$\params^*$. However, it can be argued that such a choice for the
prior is unjustified, as it can only be made {\em ex post facto}
after one has seen the data and obtained the maximum likelihood
estimate. It is more natural to appeal to a mild principle of
indifference as to the value of the parameter under the more
complex model, and thus to maximize the evidence over priors that
are symmetric about $\params^{*}$ and unimodal. This can be shown
to be equivalent to maximizing over all priors that are uniform
and symmetric about $\params^*$. This procedure leads to a very
simple expression for the lower bound on the Bayes
factor~\cite{Sellke:2001}
 \be \label{eq:Bayesian_p_vals_calibration}
 B_{10} \le \bB = { -1 \over {\rm e}\p\ln \p}
 \ee
 for $\p\le {\rm e}^{-1}$, where $\rm e$ is the exponential of
one. Here, $\p$ is the {\em p--value}, the probability that the
the value of some test statistics be as large as or larger than
the observed value {\em assuming the null hypothesis (i.e., the
simpler model) is true} (see~\xcite{Sellke:2001}{Berger:1987a} for
a detailed discussion). A more precise definition of p--values is
given in any standard statistical textbook\ct{,
e.g.~\cite{Kendall:1977}}.
\begin{table}
 \tbl{Translation table (using Eq.~\eqref{eq:Bayesian_p_vals_calibration}) between
 frequentist significance values (p--values) and
the upper bounds on the odds ($\bB$) in favour of the more complex
model. No other choice of prior (within the family considered in
the text) will give higher evidence in favour of the extra
parameters. The ``sigma'' column is the corresponding number of
standard deviations away from the mean for a normal distribution.
The ``category'' column gives the Jeffreys' scale of
Table~\ref{Tab:Jeff} (from~\cite{Gordon:2007xm}).
\label{tab:translation} }
 {\begin{tabular}{l r r r l } \toprule
  p--value &  $\bB$                 &  $\ln \bB$ &sigma&category\\
  \colrule
 0.05 & 2.5 & 0.9 & 2.0 &\\
 0.04 & 2.9 & 1.0 & 2.1 &`weak' at best\\
 0.01 & 8.0 & 2.1 & 2.6 &\\
  0.006 & 12 & 2.5 & 2.7&`moderate' at best\\
 0.003 & 21 & 3.0 & 3.0& \\
 0.001 & 53 & 4.0 & 3.3 &\\
 0.0003 & 150 & 5.0 & 3.6&`strong' at best\\
 $6\times 10^{-7}$ &
   43000 & 11 & 5.0& \\
   \botrule
   \end{tabular}}
\end{table}

Eq.~\eqref{eq:Bayesian_p_vals_calibration} offers a useful
calibration of frequentist significance values (p--values) in
terms of upper bounds on the Bayesian evidence in favour of the
extra parameters. The advantage is that the quantity on the
left--hand side of Eq.~\eqref{eq:Bayesian_p_vals_calibration} can
be straightforwardly interpreted as an upper bound on the odds for
the more complex model, whereas the p--values cannot. This point
is illustrated very clearly with an astronomical example
in~\cite{Andreon:BMIC}. In fact, a word of caution is in place
regarding the meaning of the p--value, which is often {\em
misinterpreted} as an error probability, i.e. as giving the
fraction of wrongly rejected nulls in the long run. For example,
when a frequentist test rejects the null hypothesis (in our
example, that $\params =
\params^*$) at the 5\% level, this {\em does not} mean that one
will make a mistake roughly 5\% of the time if one were to repeat
the test many times. The actual error probability is much larger
than that, and can be shown to be {\em at least} 29\% (for
unimodal, symmetric priors, see~Table 6 in~\cite{Berger:1987a}).
This important conceptual point is discussed in in greater detail
in \xcite{Berger:1987a}{Berger:2003,Berger:1987b}. The fundamental
reason for this discrepancy with intuition is that frequentist
significance tests give the probability of observing data as
extreme or more extreme than what has actually been measured, {\em
assuming} the null hypothesis ${\mathcal H}_0$ to be true (which
in Bayesian terms amounts to the choice of a model, $\mdl_0$). But
the quantity one is interested in is actually the probability of
the model $\mdl_0$ given the observations, which can only be
obtained by using Bayes' Theorem to invert the order of
conditioning\footnote{To convince oneself of the difference
between the two quantities, consider the following
example~\cite{Lyons:2006}. Imagine selecting a person at random
--- the person can either be male or female (our hypothesis). If
the person is female, her probability of being pregnant (our data)
is about $3\%$, i.e.\ $p(\rm{pregnant}|\rm{female}) = 0.03$.
However, if the person is pregnant, her probability of being
female is much larger than that, i.e.
$(\rm{female}|\rm{pregnant})\gg 0.03$. The two conditional
probabilities are related by Bayes' theorem.}. Indeed, in
frequentist statistics, a hypothesis is either true or false
(although we do not know which case it is) and it is meaningless
to attach to it a probability statement.

Table~\ref{tab:translation} lists $\bB$ for some common thresholds
for significance values and the strength of evidence scale, thus
giving a conversion table between significance values and upper
bounds on the Bayesian evidence, {\em independent of the choice of
prior for the extra parameter} (within the class of unimodal and
symmetric priors). It is apparent that in general the upper bound
on the Bayesian evidence is much more conservative than the
p--value, e.g.~a 99\% result (corresponding to $\p = 0.01$)
corresponds to odds of $8:1$ at best in favour of the extra
parameters, which fall short of even the ``moderate evidence''
threshold. Strong evidence at best requires at least a $3.6$ sigma
result. A useful rule of thumb is thus to think of a $s$ sigma
result as a $s-1$ sigma result, e.g. a $99.7\%$ result (3 sigma)
really corresponds to odds of $21:1$, i.e. about $95\%$
probability for the more complex model. Thus when considering the
detection of a new parameter, instead of reporting frequentist
significance values it is more appropriate to present the upper
bound on the Bayes factor, as this represents the maximum
probability that the extra parameter is different from its value
under the simpler model.

This approach has been applied to the cosmological context in
\cite{Gordon:2007xm}, who analysed the evidence in favour of a
non--scale invariant spectral index and of asymmetry in the cosmic
microwave background maps.  For further details on the comparison
between frequentist hypothesis testing and Bayesian model
selection, see~\cite{Ghosh:2006}.

\subsection{The effective number of parameters -- Bayesian model complexity}
\label{sec:complexity}

The usefulness of a Bayesian model selection approach based on the
Bayesian evidence is that it tells us whether the increased
``complexity'' of a model with more parameters is justified by the
data. However, it is desirable to have a more refined definition
of ``model complexity'', as the number of free parameters is not
always an adequate description of this concept. For example, if we
are trying to measure a periodic signal in a time series, we might
have a model of the data that looks like
 \begin{equation} \label{eq:timeseries}
 f(t) = A(1+ \theta \cos (t+\delta)),
 \end{equation}
where $A, \theta, \delta$ are free parameters we wish to
constrain. But if $\theta$ is very small compared to $1$ and the
noise is large compared to $\theta$, then the oscillatory term
remains unconstrained by the data and effectively we can only
measure the normalization $A$. Thus the parameters $\theta,
\delta$ should not count as free parameters as they cannot be
constrained given the data we have, and the effective model
complexity is closer to 1 than to 3. From this example it follows
that the very notion of ``free parameter'' is not absolute, but it
depends on both what our expectations are under the model, i.e.~on
the prior, and on the constraining power of the data at hand.

In order to define a more appropriate measure of complexity,
in~\cite{Spiegelhalter:2002} the notion of {\em Bayesian
complexity} was introduced, which measures the number of
parameters that the data can support. Consider the information
gain obtained when upgrading the prior to the posterior, as
measured by the the Kullback--Leibler (KL)
divergence~\ct{\cite{Kullback:1951}} between the posterior, $p$
and the prior, denoted here by $\pi$:
 \be \label{eq:def_KL}
 \KL (p,\pi) \equiv  \int p(\params|\data, \mdl)
 \ln\frac{p(\params|\data, \mdl)}{\pi(\params|\mdl)} \dr\params .
 \ee
In virtue of Bayes' theorem, $p(\params|\data, \mdl) =
\like(\params) \pi(\params | \mdl)/p(\data | \mdl)$ hence the KL
divergence becomes the sum of the negative log evidence and the
expectation value of the log--likelihood under the posterior:
 \be \label{eq:KL_derived}
 \KL (p,\pi) = - \ln p(\data | \mdl) +  \int p(\params|\data, \mdl)
 \ln \like(\params) \dr\params .
 \ee
To gain a feeling for what the KL divergence expresses, let us
compute it for a 1--dimensional case, with a Gaussian prior around
0 of variance $\Sigma^2$ and a Gaussian likelihood centered on
$\pml$ and variance $\sigma^2$. We obtain after a short
calculation
 \begin{equation}
 \KL (p,\pi) = - \frac{1}{2} - \ln \frac{\sigma}{\Sigma}
 + \frac{1}{2} \left[\left(\frac{\sigma}{\Sigma}\right)^2 \left(\frac{{\pml}^2}{\sigma^2} -1  \right)
 \right].
 \end{equation}
The second term on the right--hand side gives the reduction in
parameter space volume in going from the prior to the posterior.
For informative data, $\sigma/\Sigma \ll 1$, this terms is
positive and grows as the logarithm of the volume ratio. On the
other hand, in the same regime the third term is small unless the
maximum likelihood estimate is many standard deviations away from
what we expected under the prior, i.e. for $\pml/\sigma \gg 1$.
This means that the maximum likelihood value is ``surprising'', in
that it is far from what our prior led us to expect. Therefore we
can see that the KL divergence is a summary of the amount of
information, or ``surprise'', contained in the data.

Let us now define an effective $\chi^2$ through the likelihood as
$\like(\params) = \exp(-\chi^2/2)$. Then Eq.~\eqref{eq:KL_derived}
gives
 \be \label{eq:KL_derived_2}
 \KL (p,\pi) = - \frac{1}{2} \overline{\chi^2(\params)} +  \ln p(\data | \mdl),
 \ee
where the bar indicates a mean taken over the posterior
distribution. The posterior average of the effective chi--square
is a quantity can be easily obtained by Markov chain Monte Carlo
techniques~(see section~\ref{sec:MCMC}). We then subtract from the
``expected surprise'' the estimated surprise in the data after we
have actually fitted the model parameters, denoted by
 \be \label{eq:KL_estimator}
 \widehat{\KL} \equiv -\frac{1}{2} \chi^2(\hat{\params}) + \ln p(\data | \mdl),
 \ee
where the first term on the right--hand--side is the effective
chi--square at the estimated value of the parameters, indicated by
a hat. This will usually be the posterior mean of the parameters,
but other possible estimators are the maximum likelihood point or
the posterior median, depending on the details of the problem. We
then define the quantity
 \be \label{eq:C_b_rewritten}
 {\CC}_b \equiv -2\left(\KL(p,\pi) - \widehat{\KL} \right) = \overline{\chi^2(\params)} - \chi^2(\hat{\params}) \quad({\rm
 Bayesian~complexity}),
 \ee
(notice that the evidence term is the same in
Eqs.~\eqref{eq:KL_derived_2} and \eqref{eq:KL_estimator} as it
does not depend on the parameters and therefore it disappears from
the complexity). The Bayesian complexity gives the effective
number of parameters as {\em a measure of the constraining power
of the data as compared to the predictivity of the model},
i.e.~the prior. Hence $\CC_b$ depends both on the data and on the
prior available parameter space. This can be understood by
considering further the toy example of a Gaussian likelihood of
variance $\sigma^2$ around $\pml$ and a Gaussian prior around 0 of
variance $\Sigma^2$. Then a short calculation shows that the
Bayesian complexity is given by~(see~\cite{Kunz:2006mc} for
details)
 \be \label{eq:Cb_simple}
 {\CC}_b = \frac{1}{1 + (\sigma/\Sigma)^2}.
 \ee
So for $\sigma/\Sigma \ll 1$,  ${\CC}_b \approx 1$ and the model
has one effective, well constrained parameter. But if the
likelihood width is large compared to the prior, $\sigma/\Sigma
\gg 1$, then the experiment is not informative with respect to our
prior beliefs and ${\CC}_b \rightarrow 0$.

The Bayesian complexity can be a useful diagnostic tool in the
tricky situation where the evidence for two competing models is
about the same. Since the evidence does not penalize parameters
that are unmeasured, from the evidence alone we cannot know if we
are in the situation where the extra parameters are simply
unconstrained and hence irrelevant ($\theta \ll 1$ in the example
of Eq.~\eqref{eq:timeseries}) or if they improve the
quality--of--fit just enough to offset the Occam's razor penalty
term, hence giving the same evidence as the simpler model. The
Bayesian complexity breaks this degeneracy in the evidence
allowing to distinguish between the two cases:
\begin{enumerate}
 \item $p(\data | \mdl_0) \approx p(\data| \mdl_1)$ and $\CC_b(\mdl_1) > \CC_b(\mdl_0)$:
the quality of the data is sufficient to measure the additional
parameters of the more complicated model ($\mdl_1$), but they do
not improve its evidence by much. We should prefer model $\mdl_0$,
with less parameters.
 \item $p(\data | \mdl_0) \approx p(\data| \mdl_1)$ and $\CC_b(\mdl_1) \approx \CC_b(\mdl_0)$:
 both models have a comparable evidence and the effective
number of parameters is about the same. In this case the data is
not good enough to measure the additional parameters of the more
complicated model (given the choice of prior) and we cannot draw
any conclusions as to whether the extra parameter is needed.
\end{enumerate}
The first application of this technique in the cosmological
context is in~\cite{Kunz:2006mc}, where it is applied to the
number of effective parameters from cosmic microwave background
data\ct{, while in~\cite{Szydlowski:2008by} it was used to
determine the number of effective  dark energy parameters}.

\subsection{Information criteria for approximate model comparison}
\label{sec:information_criteria}

Sometimes it might be useful to employ methods that aim at an
approximate model selection under some simplifying assumptions
that give a default penalty term for more complex models, which
replaces the Occam's razor term coming from the different prior
volumes in the Bayesian evidence. While this is an obviously
appealing feature, on closer examination it has the drawback of
being meaningful only in fairly specific cases, which are not
always met in astrophysical or cosmological applications. In
particular, it can be argued that the Bayesian evidence (ideally
coupled with an analysis of the Bayesian complexity) ought to be
preferred for model building since it is precisely the lack of
predictivity of more complicated models, as embodied in the
physically motivated range of the prior, that ought to penalize
them.

With this {\em caveat} in mind, we list below three types of
information criteria that have been widely used in several
astrophysical and cosmological contexts. An introduction to the
information criteria geared for astrophysicists is given by
Ref.~\cite{Liddle:2004nh}. A discussion of the differences between
the different information criteria as applied to astrophysics can
be found in~\cite{Liddle:2007fy}, which also presents a few other
information criteria not discussed here.

\begin{itemize}
\item {\bf Akaike Information Criterion (AIC):} Introduced by
Akaike~\cite{Akaike:1974}, the AIC is an essentially frequentist
criterion that sets the penalty term equal to twice the number of
free parameters in the model, $k$:
  \begin{equation} \label{eq:def_AIC}
  {\rm AIC} \equiv - 2 \ln \lmax + 2k
  \end{equation}
where $\lmax \equiv p(\data | \pml, \mdl)$ is the maximum
likelihood value. The derivation of the AIC follows from an
approximate minimization of the KL divergence between the true
model distribution and the distribution being fitted to the data.

 \item {\bf Bayesian Information Criterion (BIC):}
 Sometimes called ``Schwarz Information Criterion'' (from the name
of its proposer~\cite{Schwarz:1978}), the BIC follows from a
Gaussian approximation to the Bayesian evidence in the limit of
large sample size:
  \begin{equation} \label{eq:def_BIC}
  {\rm BIC} \equiv - 2 \ln \lmax + k\ln N
  \end{equation}
where $k$ is the number of fitted parameters as before and $N$ is
the number of data points. The best model is again the one that
minimizes the BIC.
 \item {\bf Deviance Information Criterion (DIC):}
Introduced by~\cite{Spiegelhalter:2002}, the DIC can be written as
  \begin{equation} \label{eq:def_DIC}
  {\rm DIC} \equiv - 2 \widehat{\KL} + 2 \CC_b .
  \end{equation}
In this form, the DIC is reminiscent of the AIC, with the $\ln
\lmax$ term replaced by the estimated KL divergence and the number
of free parameters by the effective number of parameters, $\CC_b$,
from Eq.~\eqref{eq:C_b_rewritten}. Indeed, in the limit of
well--constrained parameters, the AIC is recovered
from~\eqref{eq:def_DIC}, but the DIC has the advantage of
accounting for unconstrained directions in parameters space.
\end{itemize}

The information criteria ought to be interpreted with care when
applied to real situations. Comparison of Eq.~\eqref{eq:def_BIC}
with Eq.~\eqref{eq:def_AIC} shows that for $N>7$ the BIC penalizes
models with more free parameters more harshly than the AIC.
Furthermore, both criteria penalize extra parameters regardless of
whether they are constrained by the data or not, unlike the
Bayesian evidence. This comes about because implicitly both
criteria assume a ``data dominated'' regime, where all free
parameters are well constrained. But in general the number of free
parameters might not be a good representation of the actual number
of effective parameters, as discussed in
section~\ref{sec:complexity}. In a Bayesian sense it therefore
appears desirable to replace the number of parameters $k$ by the
effective number of parameters as measured by the Bayesian
complexity, as in the DIC.

It is instructive to inspect briefly the derivation of the BIC.
The unnormalized posterior $g(\params) \equiv \like(\params)
p(\params|\mdl)$ can be approximated by a multi--variate Gaussian
around its mode $\sbar{\params}$, i.e. $g(\params) \approx
g(\sbar{\params}) - 1/2 (\params - \sbar{\params})^t F (\params -
\sbar{\params})$, where $F$ is minus the Hessian of the posterior
evaluated at the posterior mode. Then the evidence integral can be
computed analytically, giving
 \begin{equation}
 p(\data | \mdl) \approx \exp(g(\sbar{\params})) (2\pi)^{k/2} \vert
 F \vert ^{-1/2}.
 \end{equation}
For large samples, $N \rightarrow \infty$, the posterior mode
tends to the maximum likelihood point, $\sbar{\params} \rightarrow
\pml$, hence $g(\sbar{\params}) \rightarrow \lmax p(\pml)$ and the
log--evidence becomes
 \begin{equation} \label{eq:BIC_asympt}
 \ln p(\data | \mdl) \rightarrow \ln \lmax + \ln p(\pml) +
 \frac{k}{2}\ln(2\pi) - \frac{1}{2}\ln \vert F \vert \quad (N
 \rightarrow \infty),
 \end{equation}
where the error introduced by the various approximations scales to
leading order as $\Ord{N^{-1/2}}$. On the right--hand--side, the
first term scales as $\Ord{N}$, the second and third terms are
constant in $N$, while the last term is given by $\ln \vert F
\vert \approx k \ln N$, since the variance scales as the number of
data points, $N$. Dropping terms of order $\Ord{1}$ or
below\ct{\footnote{See~\cite{Raftery:1995b} for a more careful
treatment, where a better approximation is obtained by assuming a
weak prior which contains the same information as a single
datum.}}, we obtain
  \begin{equation} \label{eq:BICevidence}
 \ln p(\data | \mdl) \rightarrow \ln \lmax - \frac{k}{2} \ln N \quad (N
 \rightarrow \infty).
 \end{equation}
Thus maximising the evidence of Eq.~\eqref{eq:BICevidence} is
equivalent to minimizing the BIC in Eq.~\eqref{eq:def_BIC}. We see
that dropping the term $ \ln p(\pml)$ in~\eqref{eq:BIC_asympt}
means that effectively we expect to be in a regime where the model
comparison is dominated by the likelihood, and that the prior
Occam's razor effect becomes negligible. This is often not the
case in cosmology. Furthermore, this ``weak'' prior choice is
intrinsic (even though hidden from the user) in the form of the
BIC, and often it is not justified. In conclusion, it appears that
what makes the information criteria attractive, namely the absence
of an explicit prior specification, represents also their
intrinsic limitation.

\section{Cosmological parameter inference}
\label{sec:cosmo_par_inferece}

Driven by the emergence of inexpensive sensors and computing
capabilities, the amount of cosmological data has been increasing
exponentially over the last 15 years or so. For example, the first
map of cosmic microwave background (CMB) anisotropies obtained in
1992 by COBE\ncite{Smoot:1992td} contained $\sim 10^3$ pixels,
which became $\sim 5 \times 10^4$ by 2002 with
CBI~\ncite{Pearson:2002tr,Mason:2002tm}. Current state--of--the
art maps \ct{(from the WMAP satellite~\cite{Hinshaw:2006ia})}
involve $\sim 10^6$ pixels, which are set to grow to $\sim 10^7$
with Planck in the next couple of years. Similarly, angular galaxy
surveys contained $\sim 10^6$ objects in the 1970's, while by 2005
the Sloan Digital Sky Survey~\ncite{AdelmanMcCarthy:2005se} had
measured $\sim 2 \times 10^8$ objects, which will increase to
$\sim 3 \times 10^9$ by 2012 when the Large Synoptic Survey
Telescope~\ncite{Tyson:2003kb} comes online\footnote{Alex Szalay,
talk at the specialist discussion meeting ``Statistical challenges
in cosmology and astroparticle physics'', held at the Royal
Astronomical Society, London, Oct 2007.}.

This data explosion drove the adoption of more efficient map
making tools, faster component separation algorithms and parameter
inference methods that would scale more favourably with the number
of dimensions of the problem. As data sets have become larger and
more precise, so has grown the complexity of the models being used
to describe them. For example, if only 2 parameters could
meaningfully be extracted from the COBE measurement of the
large--scale CMB temperature power spectrum (namely the
normalization and the spectral
tilt\ncite{Smoot:1992td,Gorski:1994}), the number of model
parameters had grown to 11 by 2002, when smaller--scale
measurements of the acoustic peaks had become available. Nowadays,
parameter spaces of up to 20 dimensions are routinely considered.

This section gives an introduction to the broad problem of
cosmological parameter inference and highlights some of the tools
that have been introduced to tackle it, with particular emphasis
on innovative techniques. This is a vast field and any summary is
bound to be only sketchy. We give throughout references to
selected papers covering both historically important milestones
and recent major developments.

\subsection{The ``vanilla'' \LCDM~cosmological model}
\label{sec:vanilla}

Before discussing the quantities we are interested in measuring in
cosmology (the ``cosmological parameters'') and some of the
observational probes available to do so, we briefly sketch the
general framework which goes under the name of ``cosmological
concordance model''. Because it is a relatively simple scenario
containing both a cosmological constant ($\Lambda$) and cold dark
matter (CDM) (more about them below), it is also known as the
``vanilla'' \LCDM~model.

Our current cosmological picture is based on the scenario of an
expanding Universe, as implied by the observed redshift of the
spectra of distant galaxies (Hubble's law). This in turn means
that the Universe began from a hot and dense state, the initial
singularity of the Big Bang. The existence of the cosmic microwave
background lends strong support to this idea, as it is interpreted
as the relic radiation from the hotter and denser primordial era.
The expanding spacetime is described by Einstein's general
relativity. The cosmological principle states that the Universe is
isotropic (i.e., the same in all directions) and homogeneous (the
same everywhere). If follows that an isotropically expanding
Universe is described by the so--called
Friedmann--Robertson--Walker metric,
 \be
 \dr s^2 = \dr t^2 - \frac{a^2(t)}{c^2}\left[ \frac{\dr r^2}{1 +  \kappa r^2} + r^2(\dr \theta^2 + \sin^2\theta \dr \phi^2)
 \right]
 \ee
where $\kappa$ defines the geometry of spatial sections (if
$\kappa$ is positive, the geometry is spherical; if it is zero,
the geometry is flat; if it is negative, the geometry is
hyperbolic). The {\em scale factor} $a(t)$ describes the expansion
of the Universe, and it is related to redshift $z$ by
 \be
 1+z = \frac{a(t_0)}{a(t)},
 \ee
where $a(t_0)$ is the scale factor today and $a(t)$ the scale
factor at redshift $z$. The relation between redshift and comoving
distance $r$ is obtained from the above metric via the Friedmann
equation, and is given by
 \be \label{eq:cosmopars}
 a_0 \dr r = \frac{c}{H_0} \left[\Omega_\kappa (1+z)^2 + \Omega_\Lambda + (\Omega_b + \Omega_{\rm{cdm}}) (1+z)^3 + (\Omega_\gamma + \Omega_\nu) (1+z)^4
 \right]^{-1/2} \dr z,
 \ee
where $H_0$ is the {\em Hubble constant}, and $\Omega_x$ are the
{\em cosmological parameters} describing the matter--energy
content of the Universe. Standard parameters included in the
vanilla model are neutrinos ($\Omega_\nu$, with a mass $\lsim 1$
eV), photons ($\Omega_\gamma$), baryons ($\Omega_b$), cold dark
matter ($\Omega_{\rm{cdm}}$) and a cosmological constant
($\Omega_\Lambda$). The curvature term $\Omega_\kappa$ is included
for completeness but is currently not required by the standard
cosmological model (see section~\ref{sec:cosmo_model_building}).
The comoving distance determines the apparent brightness of
objects, their apparent size on the sky and the number density of
objects in a comoving volume. Hence measurements of the brightness
of standard candles, of the length of standard rulers or of the
number density of objects at a given redshift leads to the
determination of the cosmological parameters in
Eq.~\eqref{eq:cosmopars} (see next section).

The currently accepted paradigm for the generation of density
fluctuations in the early Universe is inflation.  The idea is that
quantum fluctuations in the primordial era were stretched to
cosmological scales by an initial period of exponential expansion,
called ``inflation'', possibly driven by a yet unknown scalar
field. This increased the scale factor by about 26 orders of
magnitude within about $10^{-32}$s after the Big Bang. Although
presently we have only indirect evidence for inflation, it is
commonly accepted that such a short burst of exponential growth in
the scale factor is required to solve the horizon problem, i.e. to
explain why the CMB is so highly homogeneous across the whole sky.
The quantum fluctuations also originated temperature anisotropies
in the CMB, whose study has proved to be one of the powerhouses of
precision cosmology. From the initial state with small
perturbations imprinted on a broadly uniform background,
gravitational attraction generated the complex structures we see
in the modern Universe, as indicated both by observational
evidence and highly sophisticated computer modelling.

Of course it is possible to consider completely different models,
based for example on alternative theories of gravity\ct{ (such as
Bekenstein's theory~\cite{Bourliot:2006ig,Skordis:2005xk} or
Jordan--Brans--Dicke theory~\cite{Acquaviva:2004ti})}, or on a
different way of comparing model predictions with
observations~\xcite{Wiltshire:2007jk}{Wiltshire:2007fg,Leith:2007ay}.
Discriminating among models and determining which model is in best
agreement with the data is a task for model comparison techniques,
whose application to cosmology is discussed
section~\ref{sec:cosmo_model_building}. Here we will take the
vanilla \LCDM~model as our starting point for the following
considerations on cosmological parameters and how they are
measured.

\subsection{Cosmological observations}

The discovery of temperature fluctuations in the Cosmic Microwave
Background (CMB) in 1992 by the COBE
satellite~\ncite{Smoot:1992td} marked the beginning of the era of
precision cosmology. Many other observations have contributed to
the impressive development of the field in less than 20 years. For
example, around 1990 the picture of flat Universe with both cold
dark matter and a positive cosmological constant was only
beginning to emerge, and only thanks to the painstakingly
difficult work of gluing together several fairly indirect
clues~\ncite{Efstathiou:1990xe}. At the time of writing (January
2008), the total density is known with an error of order 1\%, and
it is likely that this uncertainty will be reduced by another two
orders of magnitude in the mid--term~\ncite{Mao:2008ug}. The high
accuracy of modern precision cosmology rests on the combination of
several different probes, that constrain the physical properties
of the Universe at many different redshifts.

\begin{enumerate}
\item {\bf  Cosmic microwave background (CMB):} CMB anisotropies
offer a snapshot of the Universe at the time of recombination,
about 380,000 years after the Big Bang, at a redshift $z \approx
1100$. As described above, the temperature differences measured in
the CMB arise from quantum fluctuations during the inflationary
phase. Their usefulness lies in the fact that they are small
($\Delta T / T \sim 10^{-5}$) and hence linear perturbation theory
is mostly sufficient to predict very accurately their statistical
distribution. The 2--point correlation function of the
anisotropies is usually described via its Fourier transform, the
{\em angular power spectrum}, which presents a series of
characteristic peaks called {\em acoustic oscillations}, see
e.g.~\xcite{Challinor:2006yh}{MartinezGonzalez:2006na}. Their
structure depends in a rich way on the cosmological parameters
introduced in Eq.~\eqref{eq:cosmopars}, as well as on the initial
conditions for the perturbations emerging from the inflationary
era\ct{ (see e.g.~\cite{Hu:1995en,Hu:1995fqa,Hu:1996qs} for
further details)}. The anisotropies are polarized at the level of
1\%, and measuring accurately the information encoded by the weak
polarization signal is the goal of many ongoing observations.
\ct{State--of--the art measurements are described in
e.g.~\cite{Hinshaw:2006ia,Reichardt:2008ay,Readhead:2004gy,Montroy:2005yx,Kuo:2006ya,Jones:2005yb}.}
An example of recent measurements of the temperature power
spectrum is shown in Figure~\ref{fig:CMB}. Later this year, the
Planck satellite is expected to start full--sky, high--resolution
observations of both temperature and polarization.
\begin{figure}
\centerline{\includegraphics{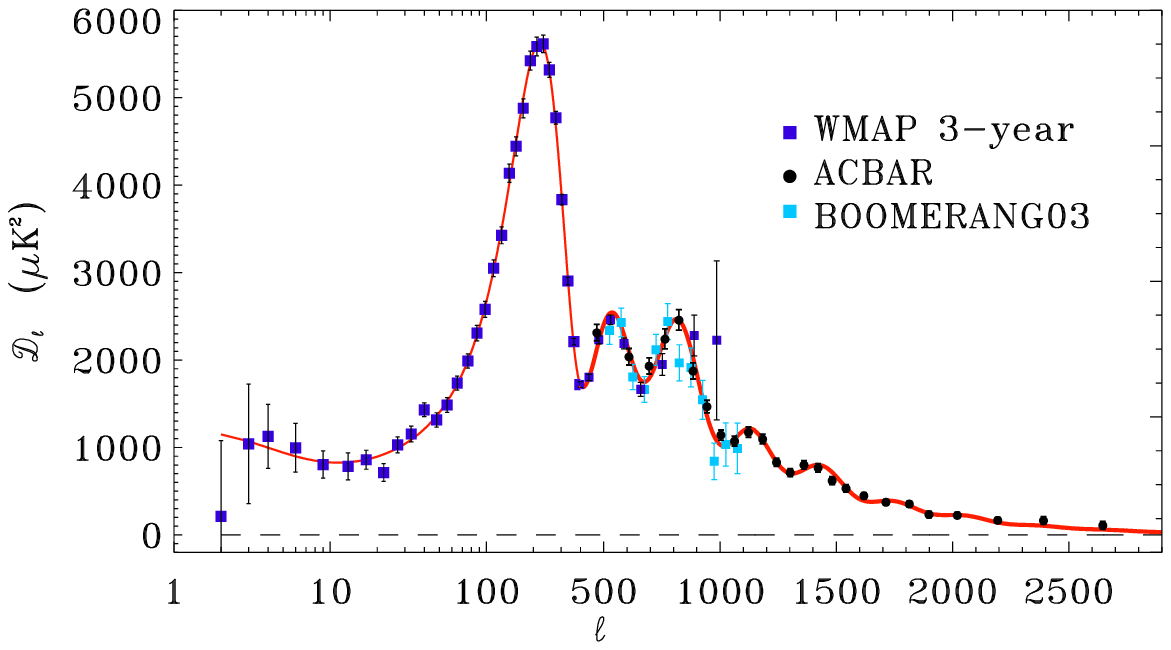}}
\caption{State--of--the-art cosmic microwave background
temperature power spectrum measurements along with the best--fit
\LCDM~model (solid line), showing data from WMAP
3--yr~\ncite{Hinshaw:2006ia}, the Boomerang 2003
flight~\ncite{Jones:2005yb} and ACBAR~\ncite{Reichardt:2008ay}
(from~\cite{Reichardt:2008ay}). \label{fig:CMB}}
\end{figure}
 \item {\bf Large scale structures (LSS):} the correlation function
among galaxies gives an estimate of the correlation properties of
the underlying dark matter distribution, up to a bias factor
relating the dark matter to the baryon distribution. Current data
typically extend out to $z\sim 0.7$. The resulting power spectrum
(recent data are shown in Figure~\ref{fig:LSS}) depends mainly on
the ratio of the radiation to matter energy density, on the
initial distribution of the perturbations with scale (spectral
index) and on the overall normalization (which can be extracted
once a bias model is specified). Heavy numerical simulation is
nowadays used to model accurately small scales, where non--linear
effects become dominant. The tool of choice to measure the power
spectrum on small scales is becoming the observations of
absorption lines from neutral hydrogen clouds, the so-called {\em
Lyman-$\alpha$ forest}~\ncite{Seljak:2006bg}, although concerns
remain about the reliability of the theoretical modelling of
non--linear effects. Recently, both the Sloan Digital Sky
Survey~\cite{Eisenstein:2005su} and the 2dF Galaxy Redshift
Survey~\cite{Cole:2005sx} have detected the presence of {\em
baryonic acoustic oscillations}, which appear as a bump in the
galaxy--galaxy correlation function corresponding to the scale of
the acoustic oscillations in the CMB. The physical meaning is that
galaxies tend to form preferentially at a separation corresponding
to the characteristic scale of inhomogeneities in the CMB.
Baryonic oscillations can be used as rulers of known length
(measured via the CMB acoustic peaks) located at a much smaller
redshift than the CMB (currently, $z\sim 0.3$), and hence they are
powerful probes of the recent expansion history of the Universe,
with particular focus on dark energy properties. The distribution
of clusters with redshift can also be employed to probe the growth
of perturbations and hence to constrain cosmology. Current galaxy
redshift surveys have catalogued about half a million objects, but
a new generation of surveys aims at taking this number to a over a
billion.
\begin{figure}
\centerline{\includegraphics[width=0.45\linewidth]{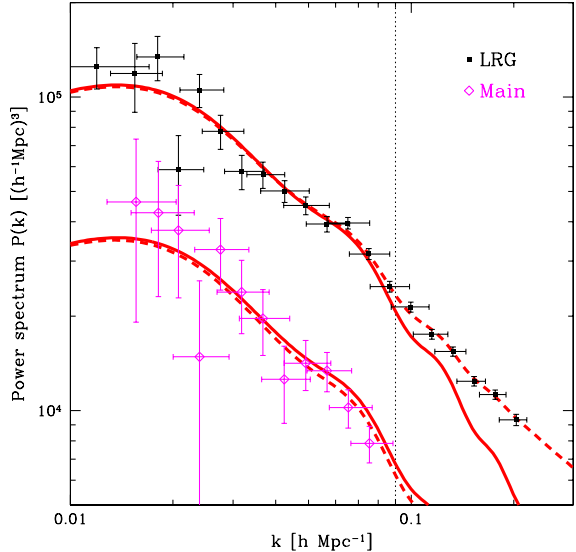}}
\caption{Current matter power spectrum measurements from the Sloan
Digital Sky Survey and best--fit \LCDM~power spectrum
(solid/dashed lines, without/with non--linear corrections),
corresponding to the cosmological parameters extracted from WMAP
3--yr CMB data (the top and bottom curves are for two different
samples). This shows that even without fitting to matter power
spectrum data, the best--fit CMB model is in good agreement with
the galaxy distribution data (from~\cite{Tegmark:2006az}).
\label{fig:LSS}}
\end{figure}

\item {\bf Supernovae type Ia:} the commonly accepted scenario for
the formation of supernovae type Ia is a white dwarf accreting
material from a binary companion. The core heats up as the
gravitational pressure increases, eventually leading to carbon
fusion ignition, followed by oxygen burning. This runaway reaction
releases within seconds a huge amount of energy, resulting in a
violent explosion which is accompanied by a massive surge in
luminosity. Observationally, type Ia supernovae are characterized
by the absence of hydrogen lines in their spectrum and they are
considered almost standard candles, in the sense that there is a
strong empirical correlation between their duration and their peak
luminosity. From measurements of their brightness as a function of
time (the light curve), their intrinsic luminosity can be
reconstructed. The data are then used to reconstruct the
redshift--distance relationship, i.e. the Hubble diagram, which in
turn depends on the cosmological
parameters~\xcite{Riess:2004nr}{Riess:2006fw,Astier:2005qq,WoodVasey:2007jb}.
Current data extend out to about $z\sim 1.4$ and encompass a few
hundreds of supernovae (a number which will increase by a factor
of 10 thanks to planned searches). Supernovae data were the first
line of evidence in 1998~\ncite{Perlmutter:1998np,Riess:1998cb}
that the expansion of the Universe is accelerating -- an effect
attributed to the existence of dark energy\ct{ (see
e.g.~\cite{Straumann:2006tv})}.

\item {\bf Weak gravitational lensing:} the presence of
inhomogeneities in the distribution of matter along the line of
sight distorts the shape of background galaxies due to the
deflection of light rays. This is most spectacularly displayed in
observations of strong lensing, showing the characteristic
arc--shaped multiple images of background galaxies. However, the
same physics affects to a much smaller degree the shape of any
background galaxy, distorting it by about 0.1 to 1\%. This is
called {\em weak gravitational lensing}. Although it is impossible
to measure such small distortions for any single galaxy, the
effect can still be detected statistically, by correlating the
shear pattern (distortion due to gravitational lensing) of several
thousand galaxies. The resulting weak lensing power spectrum
probes a combination of the geometrical setup (distance to the
background sources and to the lens) and of the parameters
controlling the growth of structures, in particular the amount of
matter (both visible and dark) and the strength of the
perturbations (for a recent review, see~\cite{Schneider:2005ka}
\ct{Some recent observations are reported
in~\cite{Hoekstra:2005cs,Jarvis:2005ck,Fu:2007qq,Benjamin:2007ys}}).
By dividing the source galaxies into slices of different redshift,
it is possible to carry out a sort of ``cosmic tomography'',
reconstructing the dark matter distribution between us and the
sources~\cite{Massey:2007wb}. Although it has not yet reached the
same level of precision of the CMB, weak lensing shows great
promise for the future in constraining cosmological parameters and
in particular dark energy.
 \end{enumerate}

\subsection{Constraining cosmological parameters}

As outlined is section~\ref{sec:inference_theory}, our inference
problem is fully specified once we give the model (which
parameters are allowed to vary and their prior distribution) and
the likelihood function for the data sets under consideration. For
the cosmological observations described above, relevant
cosmological parameters can be broadly classified in four
categories.
\begin{enumerate}
 \item {\em Parameters describing the dynamics of the background
evolution:} they represent the matter--energy content of the
Universe and its expansion history, relating redshift with
comoving distance, see Eq.~\eqref{eq:cosmopars}. The Hubble
constant today is written as $H_0 = 100h$ km/s/Mpc, and is used to
define the {\em critical energy density}, i.e. the energy density
needed to make the Universe spatially flat: $\rho_{\rm crit} =
1.88 \times 10^{-29} h^2$ g/cm$^3$. The remaining density
parameters ($\Omega_x$ in Eq.~\eqref{eq:cosmopars}) are then
written in units of the critical energy density, so that for
example the energy density in baryons is given by $\rho_b =
\Omega_b \rho_{\rm{crit}}$. Standard parameters include the energy
density in photons ($\Og$), neutrinos ($\On$), baryons ($\Ob$),
cold dark matter ($\Oc$), cosmological constant ($\Ol$) or, more
generally, a possibly time--dependent dark energy ($\Ode$).
 \item {\em Parameters describing the initial conditions for the
fluctuations:} they give the type of initial conditions, adiabatic
(where the spatial distribution of fluctuations is the same for
all fluids emerging from inflation, up to multiplicative factors)
or isocurvature (where there is a mismatch between perturbations
among two components). The most general type of initial conditions
is described by a correlation matrix that contains 10 free
parameters representing the excitation amplitude of each
mode~\xcite{Bucher:1999re}{Bucher:2004an,Trotta:2001yw}. The
simplest parameterization of the distribution of perturbations
with scale is then given for each mode in terms of a spectral
index.
 \item {\em Nuisance parameters:} these often relate
to insufficiently constrained aspects of the physics of the
observed objects, or to uncertainties in the measuring process. We
are not interested in determining them, but accounting for their
uncertainty is important in order to obtain a correct estimate of
the error on the physical parameters we are seeking to determine.
If the observable quantity has a strong dependence on poorly
determined nuisance parameters, then simply fixing the nuisance
parameters instead of marginalizing over them will lead to serious
underestimation of the uncertainty for the remaining parameters.
Example of nuisance parameters are the bias factor in galaxy
surveys, residual beam calibration uncertainty for CMB data,
supernovae intrinsic evolution parameters, intrinsic ellipticity
of background galaxies in weak lensing and others.
 \item {\em Parameters describing new physics:} this is where the
exciting frontier of data analysis lies, and we are trying to
constrain or detect effects arising from new physics in the model,
such as time--variation of the fine structure constant, the
presence of cosmic strings, the mass of neutrinos, non--trivial
topology of the Universe, extra dimensions, time--variations of
dark energy properties, and much more. Although often framed as a
parameter inference problem, this is actually a model comparison
question, and is therefore best tackled with the methods described
in section~\ref{sec:modcomp}. Therefore, in this case the
parameter inference step is only the first level towards working
out the outcome of the higher level model comparison step.
\end{enumerate}

\subsubsection{The joint likelihood function}

When the observations are independent, the log--likelihoods for
each observation simply add\footnote{This is of course not the
case when one is carrying out correlation studies, where the aim
is precisely to exploit correlation among {\em different}
observables (for example, the late Integrated Sachs--Wolfe
effect).}. Defining $\chisq \equiv -2 \ln \like$, we have that
 \begin{equation}
\label{eq:lnlike_combination} \chisq_{\rm tot} = \chisq_{\rm CMB}
+ \chisq_{\rm SN} + \chisq_{\rm lens} + \chisq_{\rm LSS} + \dots
\end{equation}
One important advantage of combining different observations as in
Eq.~\eqref{eq:lnlike_combination} is that each observable has
different {\em degenerate directions}, i.e. directions in
parameter space that are poorly constrained by the data. By
combining two or more types of observables, it is often the case
that the two data sets together have a much stronger constraining
power than each one of them separately, because they mutually
break parameters degeneracies. Combination of data sets should
never be carried out blindly, however. The danger is that the data
sets might be mutually inconsistent, in which case combining them
singles out in the posterior a region that is not favoured by any
of the two data sets separately, which is obviously
unsatisfactory. Such discrepancies might arise because of
undetected systematics, or insufficient modelling of the
observations.

In order to account for possible discrepancies of this kind,
Ref.~\cite{Lahav:1999hu} suggested to replace
Eq.~\eqref{eq:lnlike_combination} by
 \begin{equation}
 \chisq_{\rm tot} = \sum_i \alpha_i \chisq_i
\end{equation}
where $\alpha_i$ are (unknown) weight factors
(``hyperparameters'') for the various data sets, which determine
the relative importance of the observations. A non--informative
prior is specified for the hyperparameters, which are then
integrated out in a Bayesian way, obtaining an effective
chi--square
 \begin{equation}
 \chisq_{\rm eff} = \sum_i N_i \ln \chisq_i,
\end{equation}
where $N_i$ is the number of data points in data set $i$. This
method has been applied to combine different CMB observations in
the pre--WMAP era~\cite{Lahav:1999hu,Hobson:2002zf}. A technique
based on the comparison of the Bayesian evidence for different
data sets has been employed in~\cite{Marshall:2004zd}, while
Ref.~\cite{Kampakoglou:2007np} uses a technique similar in spirit
to the hyperparameter approach outlined above to perform a binning
of mutually inconsistent observations suffering from undetected
systematics, as explained in~\cite{Press:1996fw}.

After the likelihood has been specified, it remains to work out
the posterior pdf, usually obtained numerically via MCMC
technology, and report posterior constraints on the model
parameters, e.g.~by plotting 1 or 2--dimensional posterior
contours. We now sketch the way this program has been carried out
as far as cosmological parameter estimation is concerned.

\subsubsection{Likelihood--based parameter determination}

Up until around 2002, the method of choice for cosmological
parameter estimation was either direct numerical maximum
likelihood search~\xcite{Dodelson:1999am}{Melchiorri:1999br} or
evaluation of the likelihood on a grid in parameter space. Once
the likelihood has been mapped out, (frequentist) confidence
intervals for the parameters are obtained by finding the
maximum--likelihood point (or, equivalently, the minimum
chi--square) and by delimiting the region of parameter space where
the log--likelihood drops by a specified amount (details can be
found in any standard statistics textbook). If the likelihood is a
multi--normal Gaussian, then this procedure leads to the familiar
``delta chi--square'' rule--of--thumb, i.e. that e.g. a $95.4\%$
($2 \sigma$) confidence interval for 1 parameter is delimited by
the region where the $\chisq  \equiv -2 \log \like $ increases by
$\Delta \chisq = 4.00$ from its minimum value (see
e.g.~\cite{Press:2007}). Of course the value of $\Delta \chisq$
depends both on the number of parameters being constrained and on
the desired confidence interval\footnote{An important technical
point is that frequentist confidence intervals are considered
random variables --- they give the range within which our estimate
of the parameter value will fall e.g. $95.4\%$ of the time if we
repeat our measurement $N\rightarrow \infty$ times. The true value
of the parameter is given (although unknown to us) and has no
probability statement attached to it. On the contrary, Bayesian
{\em credible intervals} containing for example $95.4\%$ of the
posterior probability mass represent our degree of belief about
the value of the parameter itself. Often, Bayesian credible
intervals are imprecisely called ``confidence intervals'' (a term
that should be reserved for the frequentist quantity), thus
fostering confusion between the two. Perhaps this happens because
for Gaussian cases the two results are formally identical,
although their interpretation is profoundly different. This can
have important consequences if the true value is near the boundary
of the parameter space, in which case the results from the
frequentist and Bayesian procedure may differ substantially --
see~\cite{Host:2007wh} for an interesting example involving the
determination of neutrino masses.}

Approximate confidence intervals for each parameter were then
usually obtained from the above procedure, after maximising the
likelihood across the hidden dimensions rather than
marginalising~\xcite{Tegmark:1998nn}{Lineweaver:1998ai}, since the
latter procedure required a computationally expensive
multi--dimensional integration. The rationale was that
maximisation is approximately equivalent to marginalisation for
close--to--Gaussian problems (a simple proof can be found in
Appendix A of~\cite{Tegmark:2000db}), although it was early
recognized that this is not always a good approximation for
real--life situations~\ncite{Efstathiou:1998qr}. Marginalisation
methods based on multi--dimensional interpolation were devised and
applied in order to improve on this
respect~\xcite{Tegmark:2000db}{Tegmark:2000qy}. Many studies
adopted this methodology, which could not quite be described as
fully Bayesian yet since it was likelihood--based and the the
notion of posterior was not explicitly introduced. Often, the
choice of particular theoretical scenarios (for example, a flat
Universe or adiabatic initial conditions) or the inclusion of
external constraints (such as bounds on the baryonic density
coming from Big Bang nucleosyhntesis) were described as
``priors''. A more rigorous terminology would call the former a
model choice, while the latter amounts to inclusion (in the
likelihood) of external information. Since the likelihood could be
well approximated by a simple log--normal distribution, its
computation cost was fairly low. With the advent of
\texttt{CMBFAST}~\cite{Seljak:1996is}, the availability of a fast
numerical code for the computation of CMB and matter power spectra
meant that grids of up to 30 million points and parameter spaces
of dimensionality up to order 10 could be handled in this
way~\ncite{Tegmark:2000qy}.

\subsubsection{Bayes in the sky --- The rise of MCMC}
\label{sec:rise_MCMC}

The watershed moment after which methods based on likelihood
evaluation on a grid where definitely overcome by Bayesian MCMC
methods can perhaps be indicated in Ref.~\cite{Wang:2002rta},
which marked one of the last major studies performed using
essentially frequentist techniques. Pioneering works in using MCMC
technology for cosmological parameter extraction include the
application to VSA
data~\cite{RubinoMartin:2002rc,Slosar:2002dc}\ct{, the use of
simulated annealing~\cite{Hannestad:1999} and the study
of~Ref.~\cite{Christensen:2001gj}}. But it was the release of the
\cosmomc~code\footnote{Available from:
\texttt{cosmologist.info/cosmomc} (accessed Jan 20th 2008).} in
2002~\cite{Lewis:2002ah} that made a huge impact on the
cosmological community, as \cosmomc~quickly became a standard and
user--friendly tool for Bayesian parameter inference from CMB,
large scale structure and other data. The favourable scalability
of MCMC methods with dimensionality of the parameter space and the
easiness of marginalization were immediately recognized as major
advantages of the method. \cosmomc~employs the \texttt{CAMB}
code~\cite{Lewis:1999bs} to compute the matter and CMB power
spectra from the physical model parameters. It then employs
various MCMC algorithms to sample from the posterior distribution
given current CMB, matter power spectrum (galaxy power spectrum,
baryonic acoustic wiggles and Lyman--$\alpha$ observations) and
supernovae data.

\begin{figure}
\centerline{\includegraphics{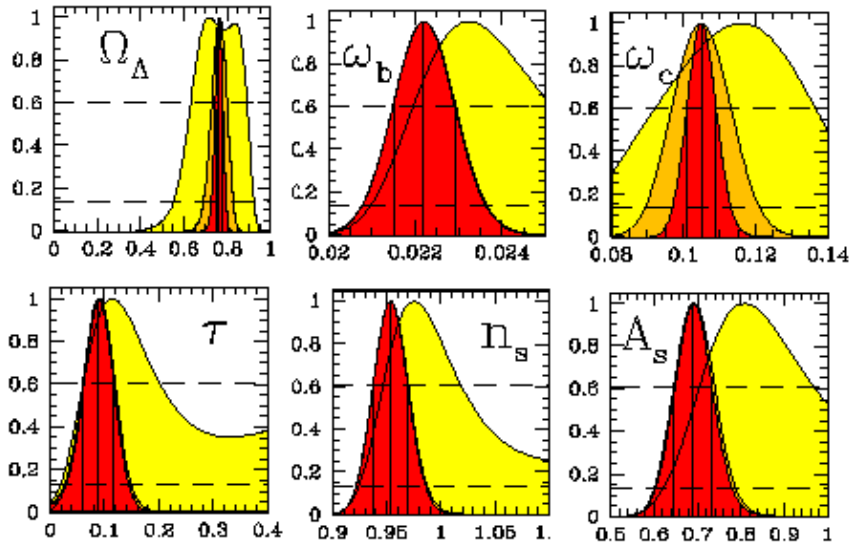}} \caption{Posterior
constraints on key cosmological parameters from recent CMB and
large scale structure data, compare Table~\ref{tab:cosmo_params}.
Top row, from left to right, posterior pdf (normalized to the
peak) for the cosmological constant density in units of the
critical density, the (physical) baryons and cold dark matter
densities. Bottom row, from left to right: optical depth to
reionization, scalar tilt and scalar fluctuations amplitude.
Yellow using WMAP 1--yr data, orange WMAP 3--yr data and red
adding Sloan Digital Sky Survey galaxy distribution data. Spatial
flatness and adiabatic initial conditions have been assumed. This
set of only 6 parameters (plus 2 other nuisance parameters not
shown here) appear currently sufficient to describe most
cosmological observations (adapted from~\cite{Tegmark:2006az}).
\label{fig:cosmo_pars}}
\end{figure}

State--of--the--art applications of cosmological parameter
inference can be found in papers such
as~\cite{Tegmark:2006az,Spergel:2006hy,Verde:2003ey,Reichardt:2008ay}.
Table~\ref{tab:cosmo_params} summarizes recent posterior credible
intervals on the parameters of the vanilla \LCDM~model introduced
above while Figure~\ref{fig:cosmo_pars} shows the full 1--D
posterior pdf for 6 relevant parameters (both from
Ref.~\cite{Tegmark:2006az}). The initial conditions emerging from
inflation are well described by one adiabatic degree of freedom
and a distribution of fluctuations that deviates slightly from
scale invariance, but which is otherwise fairly featureless.

\begin{table}
\tbl{State--of--the art cosmological parameter inference from WMAP
3--year CMB data~\cite{Hinshaw:2006ia} and Sloan Digital Sky
Survey data~\cite{Tegmark:2006az}. Posterior median and 68\%
posterior region, obtained for flat priors on the parameter set in
the top section, with the exception of the reionization optical
depth $\tau$, for which a flat prior has been adopted on
$\exp(-2\tau)$ instead (adapted
from~\cite{Tegmark:2006az}).\label{tab:cosmo_params}}
{\begin{tabular}{l l l l } \toprule
Parameter       &Value   &Meaning           &Definition\\
\colrule
\dtab{Matter budget parameters} \\
$                 \Theta_s$      &$ 0.5918^{+ 0.0020}_{- 0.0020}$                                       &CMB acoustic angular scale fit (degrees)&$\Th = r_s(\zrec)/d_A(\zrec) \times 180/\pi$\\
$                 \omega_b$      &$ 0.0222^{+ 0.0007}_{- 0.0007}$                                       &Baryon density         &$\ob=\Ob h^2 \approx \rho_b/(1.88\times 10^{-26}$kg$/$m$^3)$\\
$                 \omega_c$      &$ 0.1050^{+ 0.0041}_{- 0.0040}$                                       &Cold dark matter density       &$\ocdm=\Oc h^2 \approx \rho_c/(1.88\times 10^{-26}$kg$/$m$^3)$\\
\dtab{Initial conditions parameters} \\
$                      A_s$      &$ 0.690^{+ 0.045}_{- 0.044}$                                          &Scalar fluctuation amplitude       &Primordial scalar power at $k=0.05$/Mpc\\
$                      n_s$      &$ 0.953^{+ 0.016}_{- 0.016}$                                          &Scalar spectral index          &Primordial spectral index at $k=0.05$/Mpc\\
\dtab{Reionization history (abrupt reionization)}\\
$                     \tau$      &$ 0.087^{+ 0.028}_{- 0.030}$                                          &Reionization optical depth     &\\

\dtab{Nuisance parameters (for galaxy power spectrum)}
$                        b$      &$ 1.896^{+ 0.074}_{- 0.069}$                                          &Galaxy bias factor         & See \cite{Tegmark:2006az} for details.\\
$              Q_{\rm{nl}}$      &$30.3^{+ 4.4}_{- 4.1}$                                                &Nonlinear correction parameter & See \cite{Tegmark:2006az} for details.\\
\colrule \dtab{Derived parameters (functions of those above)}
$        \Omega_{\rm{tot}}$      &$ 1.00$ (flat Universe assumed)                                       &Total density/critical density     &$\Ot=\Om+\Ol=1-\Ok$\\
$                        h$      &$ 0.730^{+ 0.019}_{- 0.019}$                                          &Hubble parameter           &$h = \sqrt{(\ob+\ocdm)/(\Ot-\Ol)}$\\
$                 \Omega_b$      &$ 0.0416^{+ 0.0019}_{- 0.0018}$                                       &Baryon density/critical density    &$\Ob=\ob/h^2$\\
$                 \Omega_c$      &$ 0.197^{+ 0.016}_{- 0.015}$                                          &CDM density/critical density       &$\Oc=\ocdm/h^2$\\
$                 \Omega_m$      &$ 0.239^{+ 0.018}_{- 0.017}$                                          &Matter density/critical density    &$\Om=\Ob + \Oc$ \\
$           \Omega_\Lambda$      &$ 0.761^{+ 0.017}_{- 0.018}$                                          &Cosmological constant density/critical density   &$\Ol\approx h^{-2}\rho_\Lambda(1.88\times 10^{-26}$kg$/$m$^3)$\\
$                 \sigma_8$      &$ 0.756^{+ 0.035}_{- 0.035}$                                          &Density fluctuation amplitude      &See \cite{Tegmark:2006az} for details.\\

\botrule
\end{tabular}
}
\end{table}

Addition of extra parameters to this basic description (for
example, a curvature term, or a time--evolution of the
cosmological constant, in which case it is generically called
``dark energy'') is best discussed in terms of model comparison
rather than parameter inference (see next section). The breath and
range of studies aiming at constraining extra parameters is such
that it would be impossible to give even a rough sketch here.
However we can say that the simple model described by the 6
cosmological parameters given in Table~\ref{tab:cosmo_params}
appears at the moment appropriate and sufficient to explain most
of the presently available data. MCMC is nowadays almost
universally employed in one form or another in the cosmology
community.

\subsubsection{Recent developments in parameter inference}

Nowadays, the likelihood evaluation step is becoming the
bottleneck of cosmological parameter inference, as the WMAP
likelihood code~\cite{Hinshaw:2006ia} requires the evaluation of
fairly complex correlation terms, while the computational time for
the actual model prediction in terms of power spectra is
subdominant (except for spatially curved models or non--standard
scenarios containing active seeds, such as cosmic strings). This
trend is likely to become stronger with future CMB datasets, such
as Planck. Currently, for relatively straightforward extensions of
the concordance model presented in Table~\ref{tab:cosmo_params}, a
Markov chain with enough samples to give good inference can be
obtained in a couple of days of computing time on a
``off--the--shelf'' machine with 4 CPUs.

Massive savings in computational effort can be obtained by
employing neural networks techniques, a computational methodology
consistent of a network of processors called ``neurons''. Neural
networks learn in an unsupervised fashion the structure of the
computation to be performed (for example, likelihood evaluation
across the cosmological parameter space) from a training set
provided by the user. Once trained, the network becomes a fast and
efficient interpolation algorithm for new points in the parameter
space, or for classification purposes\ct{ (for example to
determine the redshift of galaxies from photometric
data~\cite{Collister:2003cz,Tagliaferri:2003})}. When applied to
the problem of cosmological parameter inference, neural networks
can teach themselves the structure of the parameter space (for
models up to about 10 dimensions) by employing as little as a few
thousands training points, for which the likelihood has to be
computed numerically as usual. Once trained, the network can then
interpolate extremely fast between samples to deliver a complete
Markov chain within a few minutes. The speed--up can thus reach a
factor of 30 or more~\xcite{Auld:2007qz}{Auld:2006pm}. Another
promising tool is a machine--learning based algorithm called
\texttt{PICO}~\xcite{Fendt:2006uh}{Fendt:2007uu}\footnote{Available
from: \texttt{http://cosmos.astro.uiuc.edu/pico/} (accessed Jan
14th 2008).}, requiring a training set of order $\sim 10^4$
samples, which are then used to perform an interpolation of the
likelihood across parameter space. This procedure can achieve a
speed--up of over a factor of 1000 with respect to conventional
MCMC.

The forefront of Bayesian parameter extraction is quickly moving
on to tackle even more ambitious problems, involving thousands of
parameters. This is the case for the Gibbs sampling technique to
extract the full posterior probability distribution for the power
spectrum $C_\ell$'s directly from CMB maps, performing component
separations (i.e., multi--frequency foregrounds removal) at the
same time and fully propagating uncertainties to the level of
cosmological parameters
~\xcite{Wandelt:2003uk,Eriksen:2007mx}{Eriksen:2004ss,Larson:2006ds}.
This method has been tested up to $\ell \lsim 50$ on WMAP
temperature data and is expected to perform well up to $\ell \lsim
100-200$ for Planck--quality data. Equally impressive is the
Hamiltonian sampling approach~\cite{Taylor:2007sj}, which returns
the $C_\ell$'s posterior pdf from a (previously foreground
subtracted) CMB map. At WMAP resolution, this involves working
with $\sim 10^5$ parameters, but the efficiency is such that the
$800$ $C_\ell$ distributions (for the temperature signal) can be
obtained in about a day of computation on a high--end desktop
computer.

Another frontier of Bayesian methods is represented by high energy
particle physics, which for historical and methodological reasons
has been so far mostly dominated by frequentist techniques.
However, the Bayesian approach to parameter inference for
supersymmetric theories is rapidly gathering momentum, due to its
superior handling of nuisance parameters, marginalization and
inclusion of all sources of uncertainty. The use of Bayesian MCMC
for supersymmetry parameter extraction has been first advocated
in~\cite{Baltz:2004aw}, and has then been rigorously applied to a
detailed analysis of the Constrained Minimal Supersymmetric
Standard
Model~\xcite{Roszkowski:2007fd,Allanach:2006cc}{Roszkowski:2007va,Trotta:2006ew,deAustri:2006pe,Allanach:2005kz},
a problem that involves order 10 free parameters. A public code
called \texttt{SuperBayeS} is available to perform an MCMC
Bayesian analysis of supersymmetric models\footnote{Code available
from: \texttt{superbayes.org} (accessed Feb 13th 2008).}.
Presently, there are hints that the constraining power of the data
is insufficient to override the prior choice in this
context\ncite{Allanach:2006jc}, but future observations, most
notably by the Tevatron or
LHC~\ncite{Roszkowski:2006mi,Allanach:2007qk} and tighter limits
(or a detection) on the neutralino scattering cross
section~\ncite{Trotta:2006ew}, should considerably improve the
situation in this respect.

\subsection{Caveats and common pitfalls}

Although Bayesian inference is quickly maturing to become an
almost automated procedure, we should not forget that a ``black
box'' approach to the problem always hides dangers and pitfalls.
Every real world problem presents its own peculiarities that
demand careful consideration and statistical inference remains
very much a craft as much as a science. While Bayes' Theorem is
never wrong, incorrect specification of the prior (for example,
making unwarranted assumptions of failing to specify relevant
external information) or inappropriate construction of the
likelihood (e.g., not reflecting the experiment or neglecting
relevant sources of uncertainty) will easily lead to wrong
inferences. Below we list a few common pitfalls that need to be
considered in the cosmological context.

\begin{enumerate}
 \item {\bf Hidden prior information.} Sometimes the choice of
flat priors on the parameters is uncritically taken to be
uninformative. This is often not the case. For instance, a flat
prior is indeed non--informative if we are estimating the mean of
a Gaussian, but if we are interested in its standard deviation
$\sigma$ a prior which is flat in $\ln \sigma$ (``Jeffreys'
prior'') is instead the appropriate choice (see
e.g.~\cite{Box:1992}). When considering extensions of the \LCDM~
model, for example, it is common practice to parameterize the new
physics with a set of quantities about which very little (if
anything at all) is known {\em a priori}. Assuming flat priors on
those parameters is often an unwarranted choice. Flat priors in
``fundamental'' parameters will in general not correspond to a
flat distribution for the observable quantities that are derived
from those parameters, nor for other quantities we might be
interested in constraining. Hence the apparently non--informative
choice for the fundamental parameters is actually highly
informative for other quantities that appear directly in the
likelihood. It is extremely important that this ``hidden prior
information'' be brought to light, or one could mistake the effect
of the prior for constraining power of the data. An effective way
to do that is to plot pdf's for the quantities of interest without
including the data, i.e. run an MCMC sampling from the prior only
(see~\cite{Slosar:2002dc} for an instructive example).
 \item {\bf No well--defined prior measure.}
Another difficulty is that in many cases several physically
equally plausible parameterizations exist, in particular for
problems involving unknown amplitudes, such as for example
isocurvature modes in the initial conditions. Since a flat prior
on parameterization $A$ is not flat in parameterization $B$ if the
two are related by a non--linear transformation, see
Eq.~\eqref{eq:prior_transformation}, two physically equivalent
setups might lead to widely different inferences. In the absence
of theoretical or physical reasons to prefer parameterization $A$
or $B$ this leads to the unsatisfactory dependence of the
posterior on the volume enclosed by the prior, as discussed for
the problem of general initial conditions in~\cite{Bucher:2004an}.
\ct{For an example of such a ``prior volume effect'' applied to
inflationary models parameters, see~\cite{Ballesteros:2007te}.}
Other obvious domains where this effect might be problematic are
dark energy equation of state reconstruction \ct{(e.g.,
\cite{Huterer:2004ch,Zunckel:2007jm})}, initial power spectrum
reconstruction~\ncite{Bridle:2003sa,Mukherjee:2005dc,Leach:2005av}
and determination of inflationary potential
parameters~\ncite{Kinney:2006qm,Peiris:2006sj,Peiris:2006ug}. In
some special cases, fundamental principles can be invoked to
define the appropriate prior measure, which exploits either
symmetries or invariance properties of the problem. An important
example is image reconstruction, where the Maximum Entropy
principle is employed to define an informative prior on the image
space (see
e.g.~\xcite{Bridle:1998ee,Marshall:2001ax}{Bridle:2000ce} for  an
astronomical application).

 \item {\bf Lindley's paradox.} A methodological issue is the
widespread use of inappropriate tools to answer what is actually a
model selection question. Often we want to assess the
``significance'' of an effect, when a deviation is observed in the
posterior from the parameter value that would correspond to the
absence of that effect. This situation is extremely common across
very different domains, from estimation of photon number counts in
presence of a background
(see~\xcite{Protassov:2002sz}{Park:2006ug,D'Agostini:2004dp} for a
proper Bayesian treatment), to the assessment of the anomalies in
the large--scale CMB power spectrum \ct{(\cite{Niarchou:2003hz}
compare frequentist methods with the Bayesian evidence)}, the
detection of gravitational waves \ct{(\cite{Allen:2002jw}
introduce a Bayesian technique that is automatically optimal)} or
the discovery of extra--solar planets~\ncite{Ford:2006}. In
general, the ``number of sigma'' away from the expected value in
the absence of a signal is {\em not} a good indicator of the
significance of the effect, a result that goes under the name of
``Lindley's paradox''~\cite{Lindley:1957}. For further details,
see Appendix A in~\cite{Trotta:2005ar}.
  \item
  {\bf Using the data twice.}
Often the data are used to guide in a qualitative fashion the
model building or the choice of priors. If the same data are then
used for a quantitative comparison the significance of the effect
can be drastically overestimated. This is a well--known problem in
frequentist statistics, which therefore insists that one should
design one's statistical tests before seeing the data at all. In
cosmology this might often be problematic or impossible to
achieve, as new observations will uncover completely unexpected
phenomena (for example, the large--scale anomalies in the cosmic
microwave background). It is however important to keep in mind
that the significance of such effects is difficult to assess.
\end{enumerate}


\section{Cosmological Bayesian model building}
\label{sec:cosmo_model_building}


The Bayesian model comparison approach based on the evaluation of
the evidence is being increasingly applied to model building
questions such as: are isocurvature contributions to the initial
conditions required by the
data~\cite{Beltran:2005xd,Trotta:2005ar,Trotta:2006ww,Lazarides:2004we}?
Is the Universe
flat~\cite{Jaffe:1995qu,Trotta:2005ar,Kunz:2006mc}? What is the
best description of the primordial power spectrum for density
perturbations~\cite{Mukherjee:2005wg,Parkinson:2006ku,Trotta:2005ar,Bridges:2006zm,Ballesteros:2007te,Bridges:2005br,Kunz:2006mc}?
Is dark energy best described as a cosmological
constant~\cite{Saini:2003wq,Bassett:2004wz,Mukherjee:2005wg,Serra:2007id,Gong:2007se,Elgaroy:2006tp,John:2002gg,Balbi:2007mz}?
In this section we review the status of the field.

\subsection{Evidence for the cosmological concordance model}

Table~\ref{tab:evidence_summary} is a fairly extensive compilation
of recent results regarding possible extensions to (or reduction
of) the vanilla \LCDM~concordance cosmological model introduced in
section~\ref{sec:vanilla}. We have chosen to compile only results
obtained using the full Bayesian evidence, rather than approximate
model comparisons obtained via the information criteria because
the latter are often {\em not} adequate approximations, for the
reasons explained in section~\ref{sec:information_criteria}. Of
course, the outcome depends on the Occam's razor effect brought
about by the prior volume (and sometimes, by the choice of
parameterization). Where applicable, we have show the sensitivity
of the result on the prior assumptions by giving a ballpark range
of values for the Bayes factor, as presented in the original
studies. The reader ought to refer to the original works for the
precise prior and parameter choices and for the justification of
the assumed prior ranges.

\begin{table}
\tbl{Summary of model comparison results against the \LCDM~
concordance model (see Table~\ref{tab:cosmo_params}) using
Bayesian model comparison for nested models. A negative (positive)
value for $\ln B$ indicates that the competing model is
disfavoured (supported) with respect to the \LCDM~model. The
column $\Delta N_{\rm{par}}$ gives the difference in the number of
free parameters with respect to the \LCDM~concordance model. A
negative value means that one of the parameters has been fixed.
See references for full details and in particular for the choice
of priors on the model parameters, which control the strengths of
the Occam's razor effect. \label{tab:evidence_summary}}
 {\begin{tabular}{l l l l l l}\toprule
 Competing model & $\Delta N_{\rm{par}}$ &  $\ln B$ & Ref & Data  & Outcome \\ \colrule
 \dtab{Initial conditions}
 Isocurvature modes & \\ \colrule
 CDM isocurvature   & $+1$ & $-7.6$ &  \cite{Trotta:2005ar}& WMAP3+, LSS
 & Strong evidence for adiabaticity \\
 + arbitrary correlations & $+4$ & $-1.0$ &
 \cite{Beltran:2005xd}& WMAP1+, LSS, SN Ia & Undecided \\
 Neutrino entropy & $+1$ & \range{-2.5}{-6.5}$^p$ & \cite{Trotta:2006ww} & WMAP3+,
 LSS& Moderate to strong evidence for adiabaticity \\
  + arbitrary correlations & $+4$ & $-1.0$ &
 \cite{Beltran:2005xd}& WMAP1+, LSS, SN Ia & Undecided \\
 Neutrino velocity & $+1$ & \range{-2.5}{-6.5}$^p$ & \cite{Trotta:2006ww} & WMAP3+,
 LSS& Moderate to strong evidence for adiabaticity \\
 + arbitrary correlations & $+4$  & $-1.0$  &
 \cite{Beltran:2005xd}& WMAP1+, LSS, SN Ia & Undecided \\

 \\  \colrule
 \dtab{Primordial power spectrum}
 No tilt ($n_s = 1$) & $-1$ &  $+0.4$ & \cite{Bridges:2005br} & WMAP1+, LSS & Undecided \\
                       &      & \range{-1.1}{-0.6}$^p$  &\cite{Mukherjee:2005wg}& WMAP1+, LSS & Undecided \\
                       &      & $-0.7$ & \cite{Trotta:2005ar} & WMAP1+, LSS &  Undecided \\
                       &      & $-0.9$   & \cite{Kunz:2006mc} & WMAP1+ & Undecided \\
                       &      & \range{-0.7}{-1.7}$^{p,d}$ & \cite{Bridges:2006zm} & WMAP3+ & $n_s = 1$ weakly disfavoured \\
                       &      & $-2.0$  & \cite{Parkinson:2006ku} & WMAP3+, LSS& $n_s = 1$ weakly disfavoured\\
                       &      & $-2.6$   & \cite{Kunz:2006mc} & WMAP3+ & $n_s = 1$ moderately disfavoured \\
                       &      & $-2.9$   & \cite{Trotta:2005ar} & WMAP3+, LSS & $n_s = 1$ moderately disfavoured\\
                       &      & $<-3.9^c$   & \cite{Gordon:2007xm} & WMAP3+, LSS & Moderate evidence at best against $n_s \neq 1$\\

 Running               & $+1$ & \range{-0.6}{1.0}$^{p,d}$ & \cite{Bridges:2006zm} & WMAP3+, LSS &
 No evidence for running
 \\
                       &       & $< 0.2^c$ & \cite{Ballesteros:2007te} & WMAP3+, LSS& Running not required\\
 Running of running    & $+2$ & $<0.4^c$  & \cite{Ballesteros:2007te} & WMAP3+, LSS & Not required
\\
 Large scales cut--off & $+2$ & \range{1.3}{2.2}$^{p,d}$ &  \cite{Bridges:2006zm} &  WMAP3+, LSS & Weak support for a cut--off
 \\ \colrule
 \dtab{Matter--energy content}
 Non--flat Universe & $+1$ & $-3.8$ &
 \cite{Kunz:2006mc} & WMAP3+, HST& Flat Universe moderately favoured \\
                    &    & $-3.4$ &
 \cite{Trotta:2005ar}& WMAP3+, LSS, HST& Flat Universe moderately favoured \\
 Coupled neutrinos & $+1$ & $-0.7$ &
 \cite{Trotta:2005ml} & WMAP3+, LSS & No evidence for non--SM neutrinos \\

  \colrule
 \dtab{Dark energy sector}
 $w(z)= w_{\rm eff} \neq -1$ & $+1$ & \range{-1.3}{-2.7}$^p$ & \cite{Saini:2003wq} & SN Ia
 &Weak to moderate support for $\Lambda$  \\
                             &      & $-3.0$ & \cite{Bassett:2004wz} & SN Ia & Moderate support for $\Lambda$  \\
                             &      & $-1.1$ & \cite{Mukherjee:2005wg} &  WMAP1+, LSS, SN Ia & Weak support for $\Lambda$ \\
                             &      & \range{-0.2}{-1}$^p$ & \cite{Serra:2007id} &  SN Ia, BAO, WMAP3 & Undecided \\
                             &      & \range{-1.6}{-2.3}$^d$ & \cite{Gong:2007se} &  SN Ia, GRB & Weak support for $\Lambda$ \\
 $w(z) = w_0 + w_1 z$ & $+2$ & \range{-1.5}{-3.4}$^p$ & \cite{Saini:2003wq} & SN Ia & Weak to moderate support for $\Lambda$ \\
                              &      & $-6.0$ & \cite{Bassett:2004wz} & SN Ia & Strong support for $\Lambda$ \\
                              &      & $-1.8$ & \cite{Serra:2007id} & SN Ia, BAO, WMAP3 & Weak support for $\Lambda$ \\
 $w(z) = w_0 + w_a(1-a)$ & $+2$ & $-1.1$ & \cite{Serra:2007id} & SN Ia, BAO, WMAP3 & Weak support for $\Lambda$ \\
                             &      & \range{-1.2}{-2.6}$^d$ & \cite{Gong:2007se} &  SN Ia, GRB & Weak to moderate support for
                             $\Lambda$ \\
 \colrule
 \dtab{Reionization history}
 No reionization ($\tau = 0$)& $-1$ & $-2.6$ &
 \cite{Kunz:2006mc} & WMAP3+, HST& $\tau \neq 0$ moderately favoured \\
 No reionization and no tilt & $-2$ & $-10.3$ &
 \cite{Kunz:2006mc} & WMAP3+, HST& Strongly disfavoured \\
 \botrule
\end{tabular}}
\tabnote{$^d$~Depending on the choice of datasets.}
\tabnote{$^p$~Depending on the choice of priors.}
\tabnote{$^c$~Upper bound using Bayesian calibrated p--values, see
section~\ref{sec:Bayesian_p_vals}.}
 \tabnote{Data sets: WMAP1+ (WMAP3+): WMAP 1st year (3--yr)
data and other CMB measurements. LSS: Large scale structures data.
SN Ia: supernovae type Ia. BAO: baryonic acoustic oscillations.
GRB: gamma ray bursts.}
\end{table}

As anticipated, the 6 parameters \LCDM~concordance model is
currently well supported by the data, as the inclusion of extra
parameters is not required by the Bayesian evidence. This is shown
by the fact that most model comparisons return either an undecided
result or they support the \LCDM~model (negative values for $\ln
B$ in Table~\ref{tab:evidence_summary}). The only exception is the
support for a cut--off on large scales in the power spectrum
reported by~\cite{Bridges:2006zm}. This is clearly driven by the
anomalies in the large scale CMB power spectrum, which in this
case are interpreted as being a reflection of a lack of power in
the primordial power spectrum. Whether such anomalies are of
cosmological origin remains however an open
question~\xcite{Copi:2005ff}{Schwarz:2004gk}. If extensions of the
model are not supported, reduction of \LCDM~to simpler models is
not viable, either: recent studies employing WMAP 3--yr data find
that a scale invariant spectrum with no spectral tilt is now
weakly to moderately
disfavoured~\cite{Parkinson:2006ku,Kunz:2006mc,Trotta:2005ar,Gordon:2007xm}.
Also, a Universe with no reionization is no longer a good
description of CMB data, and a non--zero optical depth $\tau$ is
indeed required~\cite{Kunz:2006mc}.

A few further comments about the results reported in
Table~\ref{tab:evidence_summary} are in place:
\begin{enumerate}
\item  Regarding the type of initial conditions for cosmological
perturbations, all parameter extraction studies to date (with the
exception of~\cite{Keskitalo:2006qv}) find that a purely adiabatic
mode is in agreement with observations, and constrain the
isocurvature fraction to be below about 10\% for one single
isocurvature mode at the time~\cite{Trotta:2006ww} and below about
50\% for a general mixture of
modes~\xcite{Bean:2006qz}{Beltran:2004uv}. From a model selection
perspective, this means that we expect the purely adiabatic model
to be preferred over a more complex model with a mixture of
isocurvature modes. This is indeed the case, but the result is
strongly dependent on the parameterization adopted for the
isocurvature sector, which determines the strength of the Occam's
razor effect. This is a consequence of the difficulty of coming up
with a well motivated phenomenological parameterization of the
isocurvature amplitudes\ct{, see the discussion
in~\cite{Trotta:2005ar,Lazarides:2004we}}.
 \item The shape of the primordial power spectrum
has attracted considerable attention from a model comparison
perspective
\cite{Bridges:2005br,Mukherjee:2005wg,Trotta:2005ar,Kunz:2006mc,Bridges:2006zm,Parkinson:2006ku,Gordon:2007xm,Ballesteros:2007te}.
With the exception of the large scale cut--off mentioned above,
the current consensus appears to be that a power--law distribution
of fluctuations, with power spectrum $P(k) = P_0 (k/k_0)^{n_s-1}$
with $n_s < 1$ is currently the best description. This is usually
interpreted as evidence for inflation. However, a proper model
comparison of inflationary predictions involves including the
presence of tensor modes generated by gravitational waves,
parameterized in terms of their amplitude parameter $r$. Including
this extra parameter runs into the difficulty of specifying its
prior volume, as the two obvious choices of priors flat in $r$ or
$\log r$ lead to very different model comparison
results~\cite{Parkinson:2006ku}. Another problem is that the
comparison might be ill--defined, as the simpler model with $n_s =
1$ and $r = 0$ is presumably some sort of alternative, unspecified
model without inflation that would not solve the horizon problem.
On this ground alone, unless an alternative solution to the
horizon problem is put on the table, such an alternative model
would be immediately thrown out (see the discussion
in~\cite{Ballesteros:2007te}). Finally, higher--order terms in the
Taylor expansion of the power spectrum, such as a running of the
spectral index or a running of the running, are currently not
required. This appears a robust result with respect to a wide
choice of priors and data sets.
 \item Present--day constraints on the curvature of spatial
sections are of order $|\Ok| \lsim 0.01$ (with $\Ok = 0$
corresponding to a flat, Euclidean geometry), stemming from a
combination of CMB, large scale structures and supernovae data.
Choosing a phenomenological prior of width $\Delta \Omega_\kappa =
1$ around 0 delivers a moderate support for a flat Universe versus
curved models~\cite{Kunz:2006mc,Trotta:2005ar}. However, adopting
an inflation--motivated prior instead, $\Delta \Ok \sim 10^{-5}$,
would lead to an undecided result ($\ln B = 0$) for the model
comparison, as the data are not strong enough to discriminate
between the two models in this case. This can be formalized by
considering the Bayesian model complexity for the two choices of
priors, Eq.~\eqref{eq:Cb_simple}. Noticing that $\sigma/\Sigma$ is
the ratio between the likelihood and prior widths, for a prior on
the curvature parameter of width 1, $\sigma/\Sigma \sim 10^{-2}$
and $\CC_b \approx 1$, hence the parameter has been measured. But
if we take a prior width $\sim 10^{-5}$, $\sigma/\Sigma \sim
10^{3}$ hence $\CC_b \rightarrow 0$. In the latter case, we can
see from Eq.~\eqref{eq:B01_example} that the Bayes factor between
the two models $B_{01} \rightarrow 1$ and the evidence is
inconclusive, awaiting better data.
 \item Model comparisons regarding the dark energy sector
suffer from considerably uncertainty. Clearly, the model to beat
is the cosmological constant (with equation of state parameter $w
=-1$ at all redshifts), but alternative dark energy scenarios
suffer from the fundamental difficulty of motivating physically
both the parameterization of the dark energy time dependence and
the prior volume for the extra
parameters~\cite{Efstathiou:2008ed}\ct{ (see~\cite{Uzan:2006mf}
for a review of models and~\cite{Copeland:2006wr} for on overview
of recent constraints)}. However, the semi--phenomenological
studies shown in Table~\ref{tab:evidence_summary} do agree in
deeming a cosmological constant a sufficient description of the
data. This is again a consequence of the fact that no time
evolution of the equation of state is detected in the data, hence
the strength of the support in favour of the cosmological constant
becomes a function of the available parameter space under the more
complex, alternative models. Given this result, it is interesting
to ask what level of accuracy is required before our degree of
belief in the cosmological constant is overwhelmingly larger than
for an evolving dark energy, assuming of course that future data
will not detect any significant departure from $w = -1$. To this
end, a simple classification of models has been given
in~\cite{Trotta:2006pw} in terms of their effective equation of
space parameter, $\weff$, representing the time--varying equation
of state averaged over redshift with the appropriate weighting
factor for the observable~\cite{Simpson:2006bd}. The three
categories considered are ``phantom models'' (exhibiting large,
negative values for the equation of state, $-11 \leq \weff \leq
-1$), ``fluid--like dark energy'' ($-1 \leq \weff \leq -1/3$) and
``small--departures from $\Lambda$'' models ($-1.01 \leq \weff
\leq 0.99$). Assuming a flat prior on these ranges of values for
$\weff$, consideration of the Bayes factor between each of those
models and the cosmological constant shows that gathering strong
evidence against each of the models requires an accuracy on
$\weff$ of order $\seff = 0.05$ for phantom models (which are
therefore already under pressure from current data, which have an
accuracy of order $\sim 0.1$),  $\seff = 3 \times 10^{-3}$ for
fluid--like models (about a factor of 5 better than optimistic
constraints from future observations) and $\seff = 5 \times
10^{-5}$ for small--departure models. Refinements of this approach
that employ more fundamentally--motivated priors could lead to an
analysis of the expected costs/benefits from future dark energy
observations in terms of their likely model selection outcome (we
return on this issue in section~\ref{sec:other_uses_evidence}).
\end{enumerate}

\begin{table}
\tbl{Summary of model comparison results against the \LCDM~
concordance model for some alternative (i.e., non--nested)
cosmological models. A negative (positive) value for $\ln B$
indicates that the competing model is disfavoured (supported) with
respect to \LCDM. The column $N_{\rm{par}}$ gives the number of
free parameters in the alternative model. See references for full
details about the models, priors and data used.
\label{tab:evidence_summary_alternatives} }
 {\begin{tabular}{l l l l l l}\toprule
 Competing model & $N_{\rm{par}}$ &  $\ln B$ & Ref & Data  & Outcome \\ \colrule
\dtab{Alternatives to FRW}
 Bianchi ${\rm VII}_h$   & 5 to 8 & \range{-0.9}{1.2}$^{d,p}$ &
 \cite{Bridges:2006mt}& WMAP1, WMAP3 & Weak support (at best) for Bianchi
 template \\
  & 5 to 6 & \range{-0.1}{-1.2}$^p$ &
 \cite{Bridges:2007ne}& WMAP3 & No evidence after texture
 correction \\
 \colrule
 LTB models  & 4 &
 $-3.6$ &
 \cite{GarciaBellido:2008nz}& WMAP3, BAO, SN Ia & Moderate evidence against LTB \\
 \colrule
 Fractal bubble model  & 2 &
 $0.3$ &
 \cite{Leith:2007ay}& SN Ia & Undecided \\
  \colrule
\dtab{Asymmetry in the CMB}
 Anomalous dipole & 3  & $ 1.8 $ &
 \cite{Eriksen:2007pc}& WMAP3 & Weak evidence for anomalous dipole \\
&  & $ < 2.2^c $ &
 \cite{Gordon:2007xm} & WMAP3 & Weak evidence at best  \\
 \botrule
\end{tabular}}
\tabnote{$^d$~Depending on the choice of datasets.}
\tabnote{$^p$~Depending on the choice of priors.}
\tabnote{$^c$~Upper bound using Bayesian calibrated p--values, see
section~\ref{sec:Bayesian_p_vals}.}
\end{table}

Let us now turn to models that are not nested within \LCDM
--- i.e., alternative theoretical scenarios.
Table~\ref{tab:evidence_summary_alternatives} gives some examples
of the outcome of the Bayesian model comparison with the
concordance model. As above, we restrict our considerations to
studies employing the full Bayesian evidence (there are many other
examples in the literature carrying out approximate model
comparison using information criteria instead). The model
comparison is often more difficult for non--nested models, as
priors must be specified for all of the parameters in the
alternative model (and in the \LCDM~model, as well), in order to
compute the evidence ratio. The usual {\em caveats} on prior
choice apply in this case. From
Table~\ref{tab:evidence_summary_alternatives} it appears that the
data do not seem to require fundamental changes in our underlying
theoretical model, either in the form of Bianchi templates
representing a violation of cosmic isotropy \ct{(see
also~\cite{Jaffe:2005pw})}, or as Lemaitre--Tolman--Bondi models
or fractal bubble scenarios with dressed cosmological parameters.
The anomalous dipole in the CMB temperature maps is a fine example
of Lindley's paradox. When fitting a dipolar template to the CMB
maps, the effective chi--square improves by 9 to 11 units
(depending on the details of the analysis) for 3 extra
parameters~´\cite{Eriksen:2007pc,Gordon:2006ag}, which would be
deemed a ``significant'' effect using a standard goodness--of--fit
test. However, the Bayesian evidence analysis shows that the odds
in favour of an anomalous dipole are 9 to 1 {\em at best}
(corresponding to $\ln B < 2.2$), which does not reach the
``moderate evidence at best'' threshold. Hence Bayesian model
comparison is conservative, requiring a stronger evidence before
deeming an effect to be favoured.

\subsection{Other uses of the Bayesian evidence}
\label{sec:other_uses_evidence}

Beside cosmological model building, the Bayesian evidence can be
employed in many other different ways. Here we presents two
aspects that are relevant to our topic, namely the applications to
the field of {\em multi--model inference} and {\em model selection
forecasting}.
 \begin{enumerate}
\item {\bf Multi--model inference.}
 Once we realize that there are
several possible models for our data, it becomes interesting to
present parameter inferences that take into account the model
uncertainty associated with this plurality of possibilities. In
other words, instead of just constraining parameters within each
model, we can take a step further and produce parameter inferences
that are {\em averaged over the models being considered}. Let us
suppose that we have a minimal model (in our case, \LCDM) and a
series of augmented models with extra parameters. A typical
example from cosmology is dark energy (first discussed in the
context of multi--model averaging in~\cite{Liddle:2006kn}), where
the minimal model has $w = -1$ fixed and there are several other
candidate models with a time--varying equation of state,
parameterized in terms of a number of free parameters and their
priors. Let us denote by $\params$ the cosmological parameters
common to all models. For the extended models, the
redshift--dependence of the dark energy equation of state is
described by a vector of parameters $\omega_i$ (under model
$\mdl_i$). The \LCDM~model has no free parameters for the equation
of state, hence the prior on $\omega_{\Lambda{\rm CDM}}$ is a
delta function centered on $w(z) = w_0 =-1$. Then a
straightforward application of Bayes' theorem leads to the
following posterior distribution for the parameters:
 \begin{equation} \label{eq:multimodel}
 p(\params, \omega |\data) = \sum_i p(\mdl_i|\data) p(\params, \omega | \data,
 \mdl_i),
 \end{equation}
where $p(\params, \omega | \data, \mdl_i)$ is the posterior within
each model $\mdl_i$, and it is understood that the posterior has
non--zero support only along the parameter directions $\omega_i
\subset \omega$ that are relevant for the model, and
delta--functions along all other directions. Each term is weighted
by the corresponding posterior model probability,
 \begin{equation}
 p(\mdl_i|\data) = \frac{p(\data | \mdl_i) p(\mdl_i)}{\sum_i p(\data | \mdl_i)
 p(\mdl_i)}.
 \end{equation}
The prior model probabilities $p(\mdl_i)$ are usually set equal,
but a model preference can be incorporated here if necessary. The
{\em model averaged} posterior distribution of
Eq.~\eqref{eq:multimodel} then represents the parameter
constraints obtained independently of the model choice, which has
been marginalized over. Unless one of the models is overwhelmingly
more probable than the others (in which case the model averaging
essentially disappears, as all of the weights for the other models
go to zero), the model--averaged posterior distribution can be
significantly different from the model--specific distribution. A
counter--intuitive consequence is that in the case of dark energy,
the model--averaged posterior shows {\em tighter constraints
around $w = -1$} than any of the evolving dark energy models by
itself. This comes about because \LCDM~is the preferred model and
hence much of the weight in the model--averaged posterior is
shifted to the point $w=-1$~\cite{Liddle:2006kn}. \ct{For further
details on multi--model inference, see
e.g.~\cite{Hoeting:1999,Clyde:2004}.}
 \item {\bf Model selection forecasting.}
 When considering the capabilities of future experiments, it
is common stance to predict their performance in terms of
constraints on relevant parameters, assuming a fiducial point in
parameter space as the true model (often, the current best--fit
model). While this is a useful indicator for parameter inference
tasks, many questions in cosmology fall rather in the model
comparison category. A notable example is again dark energy, where
the science driver for many future multi--million--dollar probes
is to detect possible departures from a cosmological constant,
hence to gather evidence in favour of an evolving dark energy
model. It is therefore preferable to assess the capabilities of
future experiments by their ability to answer model selection
questions.

The procedure is as follows (see~\cite{Mukherjee:2005tr} for
details and the application to dark energy scenarios). At every
point in parameter space, mock data from the future observation
are generated and the Bayes factor between the competing models is
computed, for example between an evolving dark energy and a
cosmological constant. Then one delimits in parameter space the
region where the future data would {\em not} be able to deliver a
clear model comparison verdict, for example $\vert \ln B \vert <
5$ (evidence falling short of the ``strong'' threshold). The
experiment with the smallest ``model--confusion'' volume in
parameter space is to be preferred, since it achieves the highest
discriminative power between models. An application of a related
technique to the spectral index from the Planck satellite is
presented in~\cite{Pahud:2007gi,Pahud:2006kv}.

Alternatively, we can investigate the full probability
distribution for the Bayes factor from a future observation. This
allows to make probabilistic statements regarding the outcome of a
future model comparison, and in particular to quantify the
probability that a new observation will be able to achieve a
certain level of evidence for one of the models, given current
knowledge. This technique is based on the {\em predictive
distribution} for a future observation, which gives the expected
posterior on $\params$ for an observation with experimental
capabilities described by $\df$ (this might describe sky coverage,
noise levels, target redshift, etc):
 \begin{equation} \label{eq:ppod}
 p(\params | \df, \data) = \sum_i p(\mdl_i | \data) \int \dr
 \pfid_i
 p(\params | \pfid_i, \df, \mdl_i) p(\pfid_i | \data, \mdl_i).
 \end{equation}
Here, $\data$ are the currently available observations, $p(\mdl_i
| \data)$ is the current model posterior, $p(\params | \pfid_i,
\df, \mdl_i)$ is the posterior on $\params$ from a future
observation $\df$ computed assuming $\pfid_i$ are the correct
model parameters, while each term is weighted by the present
probability that $\pfid_i$ is the true value of the parameters,
$p(\pfid_i | \data, \mdl_i)$. The sum over $i$ ensures that the
prediction averages over models, as well. From Eq.~\eqref{eq:ppod}
we can compute the corresponding probability distribution for $\ln
B$ from experiment $\df$, for example by employing MCMC techniques
(further details are given in~\cite{Trotta:2007hy}). This method
is called PPOD, for {\em predictive posterior odds distribution}
and can be useful in the context of experiment design and
optimization, when the aim is to determine which choice of $\df$
will lead to the best scientific return from the experiment, in
this case in terms of model selection capabilities \ct{(see
\cite{Bassett:2004st,Bassett:2004np,Loredo:2003nm} for a
discussion of performance optimization for parameter
constraints).} For further details on Bayes factor forecasts and
experiment design, see~\cite{Trotta:BMIC}.

\end{enumerate}


\section{Conclusions}
\label{sec:conclusion}

Bayesian probability theory offers a consistent framework to deal
with uncertainty in several different situations, from parameter
inference to model comparison, from prediction to optimization.
The notion of probability as a degree of belief is far more
general than the restricted view of probability as frequency, and
it can be applied equally well both to repeatable experiments and
to one--off situations. We have seen how Bayes' theorem is a
unique prescription to update our state of knowledge in the light
of the available data, and how the basic laws of probability can
be used to incorporate all sorts of uncertainty in our inferences,
including noise (measurement uncertainty), systematic errors
(hyper--parameters), imperfect knowledge of the system (nuisance
parameters) and modelling uncertainty (model comparison and model
averaging). The same laws can equally well be applied to the
problem of prediction, and there is considerable potential for a
systematic exploration of experiment optimization and Bayesian
decision theory (e.g., given what we know about the Universe and
our theoretical models, what are the best observations to achieve
a certain scientific goal?).

The exploration of the full potential of Bayesian methods is only
just beginning. Thanks to the increasing availability of cheap
computational power, it now becomes possible to handle problems
that were of intractable complexity until a few years ago. Markov
Chain Monte Carlo techniques are nowadays a standard inference
tool to derive parameter constraints, and many algorithms are
available to explore the posterior pdf in a variety of settings.
We have highlighted how the issue of priors --- which has
traditionally been held against Bayesian methods --- is a false
problem stemming from a misunderstanding of what Bayes' theorem
says. This is not to deny that it can be difficult in practice to
choose a prior that is a fair representation of one's degree of
belief. But we should not shy away from this task --- the fact is,
there is no inference without assumptions and a correct
application of Bayes' theorem forces us to be absolutely clear
about which assumptions we are making. It remains important to
quantify as much as possible the extent by which our priors are
influencing our results, since in many cases when working at the
cutting--edge of research we might not have the luxury of being in
a data--dominated regime.

The model comparison approach can formalize in a quantitative
manner the intuitive assessment of scientific theories, based on
the Occam's razor notion that simpler models ought to be preferred
if they offer a satisfactory explanation for the observations. The
Bayesian evidence and complexity tell us which models are
supported by the data, and what is their effective number of
parameters. Multi--model inference delivers model--averaged
parameter constraints, thus merging the two levels of parameter
inference and model comparison.

The application of Bayesian tools to cosmology and astrophysics is
blossoming. As both data sets and models become more complex, our
inference tools must acquire a corresponding level of
sophistication, as basic statistical analyses that served us well
in the past are no longer up to the task. There is little doubt
that the field of cosmostatistics will grow in importance in the
future, and Bayesian methods will have a great role to play.

\section*{Acknowledgments}

I am grateful to Stefano Andreon, Sarah Bridle, Chris Gordon,
Andrew Liddle, Nicolai Meinshausen and Joe Silk for comments on an
earlier draft and for stimulating discussions, and to Martin Kunz,
Louis Lyons, Mike Hobson and Steffen Lauritzen for many useful
conversations. This work is supported by the Royal Astronomical
Society through the Sir Norman Lockyer Fellowship, and by St
Anne's College, Oxford.


\newcommand{\bibfont}{\fontsize{10}{12}\selectfont} \newcommand{\noopsort}[1]{}
  \newcommand{\printfirst}[2]{#1} \newcommand{\singleletter}[1]{#1}
  \newcommand{\switchargs}[2]{#2#1}


{\bf \em  Roberto Trotta} is the Lockyer Fellow of the Royal
Astronomical Society at the Astrophysics Department of the
University of Oxford and a Junior Fellow of St Anne's College,
Oxford. He has worked on the problem of constraining exotic
physics with cosmic microwave background anisotropies and is
interested in several aspects of observational cosmology and
astroparticle physics, in particular in relation with dark matter
and dark energy. His recent research focuses on statistical issues
in cosmology.

\end{document}